\documentclass[aps,prd,preprint,superscriptaddress,nofootinbib,longbibliography]{revtex4-2}
\usepackage{iftex}
\ifPDFTeX
  \usepackage[T1]{fontenc}
  \usepackage[utf8]{inputenc}
  \usepackage{lmodern}
\else
  \usepackage{fontspec}
\fi
\usepackage{amsmath,amssymb,bm}
\usepackage{booktabs}
\usepackage{tikz}
\usetikzlibrary{arrows.meta,calc}
\usepackage{pgfplots}
\pgfplotsset{compat=1.18}
\usepackage[unicode,hidelinks]{hyperref}
\usepackage{xurl}
\usepackage{etoolbox}
\makeatletter\patchcmd{\frontmatter@abstract@produce}{\penalty-200\relax}{}{}{\PackageWarning{main}{abstract patch failed}}\makeatother

\newcommand{\Ds}{D_{\rm kin}}
\newcommand{\bzint}{\int_{\rm BZ}}
\newcommand{\gB}{\zeta_{\rm BZ}}
\newcommand{\glat}{g_{\rm UV}}
\newcommand{\VarB}{{\rm Var}_{\rm BZ}}
\newcommand{\CovB}{{\rm Cov}_{\rm BZ}}
\newcommand{\Pperp}{\mathcal P^\perp}
\newcommand{\iu}{\mathrm{i}}
\newcommand{\sigmabar}{{\bar\sigma}}
\pgfplotsset{every axis/.append style={width=8.6cm, height=6.2cm, tick label style={font=\small}, label style={font=\small}, legend style={font=\small}}}

\begin{document}
\global\count\footins=1000

\title{Information geometry of continuum limits in lattice field theory:\texorpdfstring{\\}{ }The Gross--Neveu model with Wilson fermions}

\author{Akio Tomiya}
\email{akio@yukawa.kyoto-u.ac.jp}
\affiliation{Department of Information and Mathematical Sciences, Tokyo Woman's Christian University, Tokyo 167-8585, Japan}
\affiliation{Department of Physics, Kyoto University, Kyoto 606-8502, Japan}
\affiliation{RIKEN Center for Computational Science, Kobe 650-0047, Japan}

\begin{abstract}
When the Boltzmann weight is nonnegative, the actions of a lattice field theory define a family of probability distributions that forms a statistical manifold, and its Fisher metric gives the number of measurements needed to distinguish two nearby actions. In the large-$N$ limit of the two-dimensional lattice Gross--Neveu model with Wilson fermions, we study the continuum limit, in which the lattice spacing $a$ is taken to zero with the mass gap $\Lambda$ fixed, through the Fisher metric of the distribution of the auxiliary field $\sigma$. The renormalization condition fixes an expectation coordinate, and dimensional transmutation becomes an exponential relation between it and a natural coordinate. The main result is that the $a\to0$ limits differ between the Fisher metric of the long-distance modes, whose momenta are fixed in physical units, and the Fisher metric of the full $P(\sigma)$. In the former, the lengths of the infinitely many irrelevant directions that change the kinetic term shrink, relative to the mass-term direction, to order $a\Lambda\log(1/(a\Lambda))$ or smaller, and when several modes are combined, only the two directions $(\Lambda,m)$ of the continuum theory remain. In the latter, the kinetic-term directions shrink at the same rate as the mass-term direction, and the difference in the regularization remains in correlations at distances of the order of the lattice spacing. An exception is the deformation that only adds a contact term to the propagator: as $a\to0$ its length goes to zero in both metrics, and it becomes indistinguishable from a change of the coupling $1/g^2$ (a lattice version of Kutasov's result). The curvature of the coupling plane at fixed kinetic term diverges as $\frac12\log^2(1/(a\Lambda))$, not as the correlation volume often assumed in thermodynamic geometry, and the critical point is infinitely far in Fisher distance from any point at finite $a$.
\end{abstract}

\maketitle

\section{Introduction}
Lattice field theory defines interacting quantum field theories nonperturbatively. The continuum theory is obtained as the limit in which the lattice spacing $a$ goes to zero, but this limit is not obtained just by setting $a$ in the action to zero. The bare couplings must be adjusted (tuned) simultaneously so that some physical quantities (for example, a mass) are kept fixed. This operation is called renormalization. In the language of theory space~\cite{WilsonKogut}, the continuum limit is a curve heading to a critical point, and along this curve the effects of the infinitely many irrelevant operators indicated by Symanzik's effective theory~\cite{Symanzik1983} disappear.

On the other hand, a Euclidean path integral on a lattice gives a probability distribution if the Boltzmann weight is real and nonnegative. Theories of bosonic fields with a real action belong to this case. In theories with fermions, the weight of the Grassmann variables is not a probability distribution, but if the determinant left after integrating out the fermions is nonnegative, the distribution of the remaining bosonic fields (for example, the gauge field of lattice QCD or the auxiliary field of this paper) is a probability distribution\footnote{Theories in which the weight becomes negative or complex, such as theories at finite density or with a topological term (the sign problem), are outside the framework of this paper.}.

The field that regards a family of probability distributions as a manifold and studies the metric and connections on it is called information geometry~\cite{AmariNagaoka,Amari2016,Fujiwara2015}. The standard metric on a family of distributions is the Fisher metric. Rao was the first to regard the Fisher information matrix as a Riemannian metric on a family of distributions~\cite{Rao1945}, Chentsov showed that this metric is determined uniquely up to a constant factor by its invariance under transformations by sufficient statistics~\cite{Chentsov1982}, and Efron introduced a statistical curvature~\cite{Efron1975}. The present framework of information geometry was built mainly by Amari~\cite{Amari1982,Amari1985}. The theory of dual connections, including the e-connection and the m-connection, was developed by Amari and Nagaoka and applied to statistical inference~\cite{AmariNagaoka}. Since the relative entropy between two nearby distributions is half the square of their distance measured with the Fisher metric, the difficulty of distinguishing two distributions from samples is determined by the Fisher metric (Stein's lemma and the Cramér--Rao inequality~\cite{CoverThomas,Rao1945}).

In lattice field theory, the couplings $g_i$ of an action $S=\sum_ig_i\sum_nO_i(n)$ are natural coordinates of an exponential family~\cite{AmariNagaoka}. In this case, therefore, the theory space of lattice field theory is a statistical manifold in the sense of information geometry, and its Fisher metric is given by the connected two-point functions of the operators in the action~\cite{Ruppeiner1995,BHM2015}. The distance between two actions measured with the Fisher metric gives the number of measurements of configurations needed to distinguish the two. Since any number of local operators can be added to a lattice action, this statistical manifold is infinite dimensional. By contrast, the continuum limit of a renormalizable theory is determined by a finite number of parameters. For the $\phi^4$ theory, it has been proven that the scaling limit of lattice models approaching the critical point is a free (Gaussian) field, in five or more dimensions~\cite{Aizenman1981,Frohlich1982} and for the nearest-neighbor Ising and $\phi^4$ models in four dimensions~\cite{AizenmanDuminilCopin2021} (triviality). This is an example in which the interaction direction is lost from the directions that remain in the continuum limit\footnote{Whether a continuum limit exists, and if it does, what theory it is, are difficult problems in themselves. Constructing the four-dimensional Yang--Mills theory and showing that it has a positive mass gap is one of the Millennium Prize Problems of the Clay Mathematics Institute~\cite{JaffeWitten2006}. For the two-dimensional O(3) nonlinear sigma model, a manuscript claiming to construct, from the nearest-neighbor lattice model, an interacting continuum limit that satisfies the Osterwalder--Schrader axioms and has a positive mass gap has been released as an output of an internal model of OpenAI (not peer reviewed)~\cite{OpenAI2026O3}. The results of this paper do not depend on the results on these problems.}. Hence one expects that in the continuum limit only a finite-dimensional part remains as distinguishable directions, and the infinitely many irrelevant directions become indistinguishable. Can the continuum limit and renormalization then be rewritten as geometry on this manifold? In this paper we address the following four questions. What operation is the continuum limit for the statistical manifold (Sec.~III)? How do the infinitely many irrelevant directions disappear, and does a finite-dimensional manifold remain for a renormalizable theory (Sec.~IV)? The continuum limit requires a divergent correlation length, that is, a phase transition of second or higher order; how is the curvature constructed from the Fisher metric related to the phase transition (Sec.~V)? Is a change of scheme a redundant transformation (a gauge transformation) on the statistical manifold that does not change the physics (Sec.~VI)?

\begin{figure*}[t]
\centering
\resizebox{\textwidth}{!}{%
\begin{tikzpicture}[font=\small, >={Stealth[length=2.0mm]}, line cap=round]
\node[anchor=west] at (-1.0,2.7) {(a)};
\draw[->, thin] (0,-2.4) -- (9.6,-2.4) node[anchor=north east] {$\theta_1$};
\draw[->, thin] (0,-2.4) -- (0,2.6) node[anchor=south] {$\theta_2$};
\foreach \k/\c/\lab in {2/black!30/{\tau=\tau_2}, 1/black!50/{\tau=\tau_1}, 0/black/{\tau=0}} {
  \pgfmathsetmacro{\cc}{-0.6-0.6*\k}
  \pgfmathsetmacro{\sh}{0.4*\k}
  \draw[\c, densely dashed, thin] (0,\cc) -- (8.55,\cc);
  \draw[\c, thick, domain=0.6:8.6, samples=70] plot (\x, {\cc+2.4*exp(-0.55*(\x-0.6))});
  \foreach \u in {1.6,3.0,4.6,6.4} \fill[\c] ({\u+\sh}, {\cc+2.4*exp(-0.55*(\u+\sh-0.6))}) circle (1.5pt);
  \node[anchor=west, \c] at (8.7,{\cc+0.03}) {$\lab$};
}
\node[anchor=east] at (0,-0.6) {$-2C_1$};
\coordinate (P) at (3.0, {-0.6+2.4*exp(-0.55*2.4)});
\draw[->, very thick] (P) -- ++(1.0, 0.15) node[anchor=west] {$v_m$};
\draw[->, very thick] (P) -- ++(0, 0.85) node[anchor=south] {$v_\Lambda$};
\draw[->, very thick, blue!75!black] (P) -- (3.4, {-1.2+2.4*exp(-0.55*2.8)}) node[anchor=north west] {$v_\delta$};
\node[anchor=north east] at (P) {X};
\draw[->] (5.0,1.0) -- (7.6,1.0) node[midway, anchor=south] {$a\to0$};
\begin{scope}[xshift=13.4cm, yshift=0.0cm]
\node[anchor=west] at (-3.4,2.7) {(b)};
\node[anchor=east, align=right] at (-1.05,1.15) {long-distance\\modes};
\node[anchor=east] at (-1.05,-1.15) {full $P(\sigma)$};
\foreach \x/\a/\b in {0/0.35/0.55, 2.3/0.35/1.1, 4.6/0.35/2.4} {
  \begin{scope}[shift={(\x,1.15)}]
    \draw[black!40, thin] (-0.9,-0.9) rectangle (0.9,0.9);
    \draw[black!25, thin] (-0.9,0) -- (0.9,0); \draw[black!25, thin] (0,-0.9) -- (0,0.9);
    \clip (-0.9,-0.9) rectangle (0.9,0.9);
    \draw[thick] (0,0) ellipse ({\a} and {\b});
  \end{scope}
}
\foreach \x/\k in {0/1.0, 2.3/1.35, 4.6/1.8} {
  \begin{scope}[shift={(\x,-1.15)}]
    \draw[black!40, thin] (-0.9,-0.9) rectangle (0.9,0.9);
    \draw[black!25, thin] (-0.9,0) -- (0.9,0); \draw[black!25, thin] (0,-0.9) -- (0,0.9);
    \clip (-0.9,-0.9) rectangle (0.9,0.9);
    \draw[thick] (0,0) ellipse ({0.22*\k} and {0.43*\k});
  \end{scope}
}
\foreach \yy in {1.15,-1.15} {
  \node[anchor=west, font=\footnotesize] at (5.52,\yy) {$v_m$};
  \node[anchor=south, font=\footnotesize, inner sep=1pt] at (4.6,{\yy+0.92}) {$v_\delta$};
}
\draw[->] (-0.9,-2.5) -- (5.5,-2.5) node[anchor=west] {$a\to0$};
\end{scope}
\end{tikzpicture}}
\caption{Overview of this paper. (a) The coupling plane $(\theta_1,\theta_2)=(1/(2g^2),m_0/g^2)$ at fixed kinetic term. The solid curves are the coexistence lines of the two vacua (first-order transition), and the dots on them are the line of constant physics, which heads to the critical point $\theta_1\to\infty$ as $a\to0$. The gray curves are the coexistence lines of the planes with the kinetic term deformed by $\tau=\tau_1$, $\tau_2$, drawn on top of each other, and $v_\delta$ is the projection of the move to such a plane. The arrows at point X are the direction $v_\Lambda$ that changes the mass gap, the mass-term direction $v_m$, and the kinetic-term direction $v_\delta$ (Sec.~II D). (b) Schematic drawing, in the $(v_m,v_\delta)$ plane, of the ellipses of the Fisher metric around point X (the range of actions that cannot be distinguished with the same number of measurements). From left to right, $a$ is decreased. The upper row is the Fisher metric of the long-distance modes, and the lower row is the Fisher metric of the full $P(\sigma)$ (values in Table~\ref{tab:main}).}
\label{fig:overview}
\end{figure*}
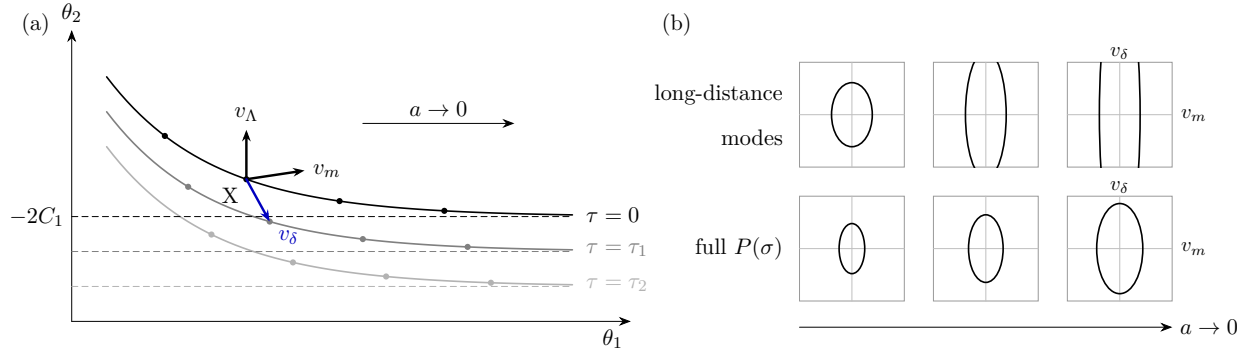

Parts of these questions have been studied before. Machta \textit{et al.} showed that the eigenvalues of the Fisher metric for irrelevant directions shrink when the observation is coarse grained~\cite{Machta2013}. Raju \textit{et al.} studied the flow of the metric under coarse graining~\cite{Raju2018}. Balasubramanian \textit{et al.} studied the relative entropy between field theories and stated that a change of scheme is a change of coordinates of the couplings~\cite{BHM2015}. Bény and Osborne defined relevant and irrelevant through equivalence classes of distributions that cannot be distinguished at a given resolution~\cite{BenyOsborneNJP}. For families of conformal field theories, the metric is given by two-point functions and the connection by contact terms, and a change of contact terms is a change of coordinates~\cite{Kutasov1989}. Pagani and Sonoda stated that there is no standard metric on theory space~\cite{PaganiSonoda2018}. The relation between phase transitions and curvature has been studied in thermodynamic geometry, and there is a view that near a critical point the curvature is proportional to the correlation volume $\xi^d$~\cite{Ruppeiner1995,JJK2003,Dolan1998,EGJ2020}. For quantum phase transitions, the critical scaling of the fidelity susceptibility (quantum information metric) of the ground state has been studied~\cite{ZanardiPaunkovic2006,CamposVenutiZanardi2007}.

In this paper, we study the continuum limit and renormalization from the viewpoint of information geometry by analyzing the two-dimensional lattice Gross--Neveu model~\cite{GrossNeveu1974} with Wilson fermions in the large-$N$ limit. The Gross--Neveu model is asymptotically free, and the fermion mass is generated dynamically through spontaneous breaking of chiral symmetry; it has therefore been studied as a toy model of QCD~\cite{GrossNeveu1974}. Moreover, it is solvable at large $N$. In addition, since the Wilson term breaks chiral symmetry explicitly, tuning of the bare mass is needed, as in lattice QCD~\cite{Aoki1984,AokiHigashijima1986}.

In the Gross--Neveu model, the Boltzmann weight of the fermions is a Grassmann number and is not a probability distribution, so we take the distribution of the auxiliary field $\sigma$ as the statistical manifold. As $N\to\infty$, the Fisher metric of this distribution splits exactly into a sum of contributions of the momentum modes of $\sigma$. Hence the contributions of physical-scale modes and lattice-scale modes can be separated without coarse graining. We call the modes whose momentum $k$ is fixed in physical units (in units of $\Lambda$) the long-distance modes (here long distance means long compared with the lattice spacing, and modes with $k\gg\Lambda$ are included), and we call the $a\to0$ limit of the Fisher metric of their marginal distribution the Fisher metric of the long-distance modes. This is the information obtained by looking only at the long-distance modes in the data of $P(\sigma)$, and it is obtained by measuring long-distance correlations. On the other hand, the Fisher metric of the full $P(\sigma)$ at lattice spacing $a$ is the information obtained by measuring all correlations down to distances of the order of the lattice spacing. Both metrics measure the lengths of the same directions with the renormalization conditions imposed. The main result of this paper is that the $a\to0$ limits differ between the two, that is, that ``fixing the modes'' and ``$a\to0$'' do not commute (Fig.~\ref{fig:overview}, Table~\ref{tab:main}).

We summarize the answers to the four questions. The continuum limit is a curve in the coupling plane at fixed kinetic term that heads to the critical point at infinity. In terms of the two dual coordinate systems, the couplings (natural coordinates) and the expectation values (expectation coordinates), the renormalization condition is a condition that fixes the long-distance expectation coordinate $\langle\sigma_n\rangle$, and dimensional transmutation becomes the exponential relation $\langle\sigma_n\rangle=e^be^{-2\pi\theta_1}$ ($\theta_1=1/(2g^2)$, and $b$ is a constant determined by the kinetic term) between a natural coordinate and an expectation coordinate (Sec.~III). The lengths of the irrelevant kinetic-term directions all go to zero in the Fisher metric of the long-distance modes, so that only the two directions $(\Lambda,m)$ remain, whereas in the Fisher metric of the full $P(\sigma)$ these directions remain, except for the contact-term direction (Sec.~IV). The curvature of the coupling plane diverges as $\frac12\ell^2$ ($\ell=\log(1/(a\Lambda))$) and grows exponentially with the Fisher distance traveled toward the critical point, so that the relation $|R|\sim\xi^d$ of thermodynamic geometry does not hold (Sec.~V). Among changes of scheme, a change of the renormalization condition is a change of coordinates, the contact-term direction becomes indistinguishable from a change of coordinates as $a\to0$, and the other changes of the kinetic term are directions whose lengths go to zero in the Fisher metric of the long-distance modes (Sec.~VI).

In Sec.~II we define the model and the statistical manifold and fix the order of limits. In Secs.~III--VI we answer the four questions in turn, in Sec.~VII we discuss the relation to previous work and the limitations of the results, and in Sec.~VIII we give a summary and future problems. Derivations are given in Appendices A--F.

\section{Model and statistical manifold}
\subsection{Lattice Gross--Neveu model}
We consider $N$ species ($N$ even) of two-component fermions on a two-dimensional square lattice. We take the lattice spacing as the unit, label the sites by $n\in\mathbb Z^2$, and take the action to be
\begin{equation}
S=\sum_n\Big[\sum_{i=1}^N\bar\psi^i_n\big((\Ds+m_0)\psi^i\big)_n-\frac{g^2}{2N}\Big(\sum_i\bar\psi^i_n\psi^i_n\Big)^2\Big],\qquad \Ds(p)=\iu\sum_\mu\gamma_\mu\kappa_\mu(p)+w(p).
\label{eq:action}
\end{equation}
For Wilson fermions (with the coefficient of the Wilson term set to 1), $\kappa_\mu=\sin p_\mu$ and $w=w_W\equiv\sum_\mu(1-\cos p_\mu)$. When $m_0=0$ and $w=0$, the action is invariant under the discrete chiral transformation $\psi\to\gamma_5\psi$, $\bar\psi\to-\bar\psi\gamma_5$, and the Wilson term $w$ breaks this symmetry explicitly. The phase structure of the lattice Gross--Neveu model with Wilson fermions has been studied, and a parity-broken phase has been found~\cite{Aoki1984,Izubuchi1998}. Since the interaction of the model in this paper is only $(\bar\psi\psi)^2$, the large-$N$ saddle point is determined by the real auxiliary field $\sigma$ (Sec.~II B) alone.

The kinetic term on the lattice, viewed in the continuum theory, contains infinitely many irrelevant operators (Symanzik's effective theory~\cite{Symanzik1983}). To study these irrelevant directions of the kinetic term, in what follows we deform the kinetic term from the Wilson form by a local (analytic on the Brillouin zone) and cubic-symmetric deformation
\begin{equation}
\begin{pmatrix}\kappa\\ w\end{pmatrix}
\to
\begin{pmatrix}\kappa\\ w\end{pmatrix}
+\tau
\begin{pmatrix}\delta\kappa\\ \delta w\end{pmatrix}.
\label{eq:deform}
\end{equation}
In what follows we call $\delta=(\delta\kappa,\delta w)^\top$ the direction of the deformation and $\tau$ its size. We assume that $\delta w$ starts at $p^2$ and $\delta\kappa$ starts at $p^3$. Hence, for any deformation, the coefficient of the linear term in $p$ of $\kappa_\mu$ stays 1, and the normalization of the kinetic term in the continuum limit does not change. The estimates below hold uniformly in the deformation (the way to measure the size of a deformation is given in Appendix A). For figures and numerical examples we use the following three directions.
\begin{equation}
{\rm d1}:\
\begin{cases}
\delta \kappa=0,\\
\delta w=w_W,
\end{cases}
\qquad
{\rm d2}:\
\begin{cases}
\delta \kappa=0,\\
\delta w=-\tfrac12\sum_\mu\hat p_\mu^4,
\end{cases}
\qquad
{\rm d3}:\
\begin{cases}
\delta\kappa_\mu=\sin p_\mu(1-\cos p_\mu),\\
\delta w=0
\end{cases}.
\label{eq:dirs}
\end{equation}
Here $\hat p_\mu=2\sin(p_\mu/2)$. d1 is the direction that changes the coefficient of the Wilson term, d2 is a direction of $w$ that changes the coefficient of $p^4$ without changing that of $p^2$, and d3 is the direction that changes the coefficient of $p^3$ of $\kappa$.

\subsection{Distribution of the auxiliary field}
Introducing the auxiliary field $\sigma_n$ and integrating out the fermions, the distribution of $\sigma$ becomes
\begin{equation}
P(\sigma)=\frac1{Z(\theta)}\det(\Ds+\sigma)^N\exp\Big\{-N\sum_n\big[\theta_1\sigma_n^2-\theta_2\sigma_n\big]\Big\},\qquad\theta_1=\frac1{2g^2},\quad\theta_2=\frac{m_0}{g^2}.
\label{eq:P}
\end{equation}
Here $\sigma=\{\sigma_n\}$ is the configuration that collects the auxiliary field on all sites, $\sigma_n$ is its value at site $n$, and the $\sigma$ in $\det(\Ds+\sigma)$ is the matrix with $\sigma_n$ on the diagonal. $Z(\theta)$ is the normalization ($\int\mathcal D\sigma\,P(\sigma)=1$, $\mathcal D\sigma=\prod_nd\sigma_n$) and is the partition function of the distribution of $\sigma$. Since the Wilson Dirac operator satisfies $\gamma_5\Ds\gamma_5=\Ds^\dagger$, $\det(\Ds+\sigma)$ is real, and $P(\sigma)\ge0$ for even $N$.

We read Eq.~(\ref{eq:P}) in the form of an exponential family. In general, when a family of distributions of a random variable $\sigma$ can be written in the form
\begin{equation*}
P(\sigma)=\frac1{Z(\theta)}h(\sigma)\exp\Big[\sum_i\theta_iT_i(\sigma)\Big],
\end{equation*}
it is called an exponential family~\cite{AmariNagaoka}. $h(\sigma)$ is a weight independent of $\theta$ (the base measure), and $T_i(\sigma)$ are sufficient statistics (quantities such that, to estimate $\theta$, it suffices to know only the values $T_i(\sigma)$ of the configuration $\sigma$). The $\theta_i$ are called natural coordinates (e-coordinates), and the expectation values $\eta_i=\langle T_i\rangle=\partial\log Z/\partial\theta_i$ are called expectation coordinates (m-coordinates). The Fisher metric is $\partial_i\partial_j\log Z=\langle T_iT_j\rangle-\langle T_i\rangle\langle T_j\rangle$. In the language of field theory, $\theta_i$ are couplings, $T_i$ are the operators that multiply them, $\log Z$ is the logarithm of the partition function (minus the free energy), and the Fisher metric is the connected two-point function of the operators. In Eq.~(\ref{eq:P}), $h=\det(\Ds+\sigma)^N$, $T_1=-N\sum_n\sigma_n^2$, and $T_2=N\sum_n\sigma_n$, so $P(\sigma)$ at fixed kinetic term is, for any $N$, an exponential family with natural coordinates $(\theta_1,\theta_2)$. By translation invariance, $\langle\sigma_n^2\rangle$ and $\langle\sigma_n\rangle$ do not depend on $n$, and up to constant factors $(\langle\sigma_n^2\rangle,\langle\sigma_n\rangle)$ are the expectation coordinates. The size $\tau$ of a deformation of the kinetic term enters the base measure $h$, and since $\partial_\tau^2\log P$ depends on $\sigma$, $\tau$ is not a natural coordinate. If one directly takes the family of fermionic Boltzmann weights $e^{-S}$, the second derivative of the logarithm of the partition function becomes negative in kinetic-term directions and does not give a metric (Appendix F). This is why we take the distribution of $\sigma$ as the statistical manifold.

\subsection{\texorpdfstring{Order of limits and the large-$N$ theory}{Order of limits and the large-N theory}}
\label{sec:limits}
In this paper, we take the limits in the order $N\to\infty$, infinite volume, and $a\to0$. As $N\to\infty$, $\sigma$ becomes Gaussian around the saddle point $\sigma_c$. The two-point function of the Fourier components $\delta\sigma_k$ of the fluctuation $\delta\sigma_n=\sigma_n-\sigma_c$ is
\begin{equation}
\langle\delta\sigma_k\delta\sigma_{-k}\rangle=\frac{G(q)}N,\qquad G^{-1}(q)=\frac1{g^2}+\Pi(q),\qquad q=ak.
\label{eq:G}
\end{equation}
$q$ is the momentum in lattice units, and $k$ is the momentum in physical units. $\Pi(q)$ is the self-energy of $\sigma$ from the fermion one-loop, given at the saddle-point value $\sigma_c$ by
\begin{equation}
\Pi(q)=\bzint\frac{2[M(p)M(p+q)-\kappa(p)\cdot\kappa(p+q)]}{D(p)D(p+q)},\qquad M=\sigma_c+w,\quad D=\kappa^2+M^2.
\label{eq:Pi}
\end{equation}
$D(p)$ is an eigenvalue of $(\Ds+\sigma_c)^\dagger(\Ds+\sigma_c)$. Here $\bzint f=\int_{[-\pi,\pi]^2}f\,d^2p/(2\pi)^2$ is the average of $f$ over the Brillouin zone (we use the same symbol when the integration variable is $q$). We write the variance and covariance over the Brillouin zone as $\VarB(f)=\bzint f^2-\big(\bzint f\big)^2$ and $\CovB(f,f')=\bzint ff'-\bzint f\bzint f'$. $\langle\cdots\rangle$ is used only for expectation values in the distribution of $\sigma$. We call the theory in which $P(\sigma)$ is replaced by this Gaussian distribution the large-$N$ theory, and all results below refer to this distribution. We write the average of the inverse propagator over the Brillouin zone as $1/\glat^2\equiv\bzint G^{-1}$. $\glat^2$ is the size of the propagator of $\sigma$ at lattice-scale momenta and represents the effective coupling at the lattice scale (Sec.~IV A).

The saddle point $\sigma_c$ is the value of $\sigmabar$ that minimizes the effective potential per site at the uniform configuration $\sigma_n=\sigmabar$ ($\sigmabar$ a constant), $V(\sigmabar)=\theta_1\sigmabar^2-\theta_2\sigmabar-I(\sigmabar)$, $I(\sigmabar)=\bzint\log[\kappa^2+(\sigmabar+w)^2]$. It is a function of the couplings and the kinetic term, and it agrees with the expectation coordinate $\langle\sigma_n\rangle$ up to $1/N$ (Sec.~III). At fixed $\theta_1$ and kinetic term, $\sigma_c$ and $\theta_2$ determine each other through the stationarity condition $V'(\sigma_c)=0$, that is, $\theta_2=2\theta_1\sigma_c-I'(\sigma_c)$, and we also use $\sigma_c$ as a coordinate in place of $\theta_2$. In the $(\theta_1,\theta_2)$ plane at fixed kinetic term, there is a coexistence line (first-order transition) on which the two minima $\sigma_\pm$ are equally deep (Fig.~\ref{fig:overview}(a)). What is fixed in the continuum limit is $\langle\sigma_n\rangle=\sigma_+=a\Lambda$ in the vacuum on the positive side and the coexistence of the two vacua. $\Lambda$ is the mass gap of the fermion (the dynamically generated mass). These two conditions determine the line of constant physics in the chiral limit. In what follows we write $\ell=\log(1/(a\Lambda))$. $\ell$ is the logarithm of the ratio of the cutoff $1/a$ to the mass gap $\Lambda$, and $a\to0$ corresponds to $\ell\to\infty$.

On the coexistence line, the distribution of $\sigma$ in a finite volume is a mixture of the distributions around the two vacua. The Fisher metric of the mixture has, in the direction across the line, a term coming from the change of the weights of the two vacua, and this term diverges in proportion to the number of sites. In this paper, we take $N\to\infty$ first and adopt the Gaussian distribution around the vacuum $\sigma_+$ on the positive side as the definition. This corresponds to the distribution obtained by shifting slightly off the coexistence line to the positive side, taking the infinite-volume limit, and then returning to the line, that is, the distribution restricted to one of the vacua.

\subsection{Fisher metric and mode decomposition}
\label{sec:metric}
The Fisher metric (per site) of the large-$N$ theory is
\begin{equation}
ds^2=NG^{-1}(0)\,d\sigma_c^2+\frac12\bzint(d\log G)^2.
\label{eq:metric}
\end{equation}
$d$ denotes the differential when $\sigma_c$, $\theta_1$, and the size $\tau$ of the deformation of the kinetic term (one for each direction $\delta$ of deformation) are varied. The first term is the information of the mean of $\sigma$ (the uniform component) and has rank 1. The second term is the information of the covariance and splits exactly into a sum over the momentum modes of $\sigma$. The contribution of each mode is obtained by reading off one term of this sum.

The Fisher metric has the following meaning. The relative entropy between two nearby distributions whose coordinates differ by $d\theta$ is $ds^2/2$ per site. In a test that distinguishes two actions by measuring $N_{\rm conf}$ configurations of $\sigma$ on a lattice with $N_s$ sites, the error probability decreases as $\exp(-N_{\rm conf}N_s\,ds^2/2)$ (Stein's lemma~\cite{CoverThomas}). Hence two actions can be distinguished roughly when $N_{\rm conf}N_s\,ds^2\gtrsim1$. The same magnitude is obtained from the Cramér--Rao inequality~\cite{Rao1945} for the error in estimating the coordinates. The contribution $\frac12(d\log G(q))^2$ of mode $q$ is the information obtained by measuring the correlation $\langle\delta\sigma_k\delta\sigma_{-k}\rangle$ of that mode. Hence a direction whose length in the Fisher metric of the long-distance modes goes to zero as $a\to0$ is a difference of actions that becomes harder to distinguish, with a fixed number of measurements of long-distance correlations, as $a$ is decreased. To distinguish a kinetic-term direction with long-distance correlations, $1/r_\delta$ times as many measurements (of order $1/(a\Lambda\ell)^2$ times as many for the direction that changes the coefficient of the Wilson term) are needed as for a difference of the same size in the mass-term direction (Sec.~IV B). A direction that remains in the Fisher metric of the full $P(\sigma)$ is a difference that can be distinguished by measuring correlations at distances of the order of the lattice spacing.

We use the following three directions to move on the manifold. The first is the direction $v_\Lambda=\partial_{\sigma_c}|_{\theta_1}$ that changes $\sigma_c$ with $\theta_1$ fixed; since $\sigma_c=a\Lambda$, it corresponds to changing the mass gap $\Lambda$ with $a$ fixed. The second is the direction $v_m=\frac12\partial_{\theta_1}|_{\sigma_c}$ that changes $\theta_1$ with $\sigma_c$ fixed. Since $\Pi$ does not change when $\sigma_c$ is fixed, Eq.~(\ref{eq:G}) shows that $G^{-1}$ shifts by the same amount at all $q$, and $v_mG^{-1}=1$. Moving in the direction $v_m$ leaves the coexistence line and lifts the degeneracy of the two vacua $\sigma_\pm$. In the continuum theory, this corresponds to adding a mass term $m$ for the fermion, which explicitly breaks the discrete chiral symmetry, so we call $v_m$ the mass-term direction. The third is the kinetic-term direction $v_\delta$, the tangent vector that deforms the kinetic term in the direction $\delta$ and readjusts $m_0$ and $g^2$ so as to stay on the line of constant physics. Since $\sigma_c$ does not move along $v_m$ and $v_\delta$, the first term of Eq.~(\ref{eq:metric}) does not contribute. Since $v_\Lambda$ has a squared length $NG^{-1}(0)$ proportional to $N$ from the first term, we take $v_m$, which likewise has only the second term, as the reference with which the kinetic-term directions are compared.

We measure how distinguishable a kinetic-term direction is, compared with the mass-term direction, by the ratio for each mode $q$,
\begin{equation}
r_\delta(q)=\frac{\frac12G(q)^2[v_\delta G^{-1}(q)]^2}{\frac12G(q)^2[v_mG^{-1}(q)]^2}=[v_\delta G^{-1}(q)]^2.
\label{eq:ratio}
\end{equation}
Since the factors $G^2$ in the numerator and the denominator cancel, $r_\delta$ does not depend on the normalization of the metric. For the Fisher metric of the full $P(\sigma)$ we use the second term of Eq.~(\ref{eq:metric}), and we write the component for directions $v$ and $v'$ as $F_{vv'}=\frac12\bzint G^2(vG^{-1})(v'G^{-1})$. Here $vG^{-1}$ is the derivative of $G^{-1}$ in the direction $v$. We write the squared length of the component of $v_\delta$ that is Fisher orthogonal to $v_m$ (the component whose inner product with $v_m$ measured with the Fisher metric vanishes) as $F^\perp_\delta=F_{\delta\delta}-F_{m\delta}^2/F_{mm}$. The main results are summarized in Table~\ref{tab:main}.

\begin{table}[t]
\centering
{\small
\begin{tabular}{l@{\hspace{1.2em}}l@{\hspace{1.2em}}l}
\toprule
 & \shortstack[l]{Fisher metric of the\\long-distance modes} & \shortstack[l]{Fisher metric of\\the full $P(\sigma)$}\\
\midrule
Squared length of $v_m$ & $\frac12G_c(k/\Lambda)^2$ & $\pi^2/(2\ell^2)$\\
\shortstack[l]{Ratio of kinetic-term\\direction to $v_m$} & $(a\Lambda)^2\ell^2$ (d1), $(a\Lambda)^2$ (d2) $\to0$ & finite (0.264 for d1, 0.665 for d2)\\
\shortstack[l]{Directions with\\nonzero length} & $v_\Lambda$ and $v_m$ (several modes) & \shortstack[l]{$v_m$ and kinetic-term directions\\(except the contact term)}\\
\shortstack[l]{Information on\\kinetic-term directions} & vanishes as $a\to0$ & \shortstack[l]{correlations at one or two\\lattice spacings}\\
\bottomrule
\end{tabular}}
\caption{The $a\to0$ limits in the two metrics. The Fisher metric of the long-distance modes is the metric of the modes with momentum $k$ fixed in physical units. Since the first term of Eq.~(\ref{eq:metric}) does not contribute for $v_m$ (the mass-term direction) and $v_\delta$, the Fisher metric of the full $P(\sigma)$ coincides with the covariance part (the second term). The length of the contact-term direction $\delta\Ds\propto\Ds^2$ goes to zero in both metrics as $a\to0$, and in the eight-dimensional space of directions examined numerically, it is the only direction that is degenerate in the Fisher metric of the full $P(\sigma)$ (Sec.~IV D). $v_\Lambda$ separately remains finite through the mean part $NG^{-1}(0)\to N/\pi$.}
\label{tab:main}
\end{table}

\section{Renormalization condition and tuning}
\label{sec:tuning}
The meaning of the renormalization condition can be read off immediately from the structure of the exponential family in Sec.~II B. $\langle\sigma_n\rangle$ in the vacuum on the positive side is an expectation coordinate, dual to the natural coordinate $\theta_2=m_0/g^2$. The renormalization condition $\langle\sigma_n\rangle=a\Lambda$ is a condition that fixes this long-distance expectation coordinate. The other expectation coordinate $\langle\sigma_n^2\rangle$ is a short-distance quantity, since $\langle\sigma_n^2\rangle-\langle\sigma_n\rangle^2=N^{-1}\bzint G$ is determined by lattice-scale modes. The coexistence condition expresses that we are at a kink of the logarithm of the partition function, $\log Z$. The mixed coordinates $(\theta_1,\langle\sigma_n\rangle)$, which pair one natural coordinate with one expectation coordinate, are Fisher orthogonal~\cite{AmariNagaoka}. Exact orthogonality holds when $\langle\sigma_n\rangle$ is taken as a coordinate; when the saddle point $\sigma_c$ is taken as a coordinate as in Sec.~II D, a cross term between $v_\Lambda$ and $v_m$ appears in the $N^0$ part of the metric because of the $1/N$ difference between the two.

Solving the conditions for the line of constant physics, the tuning in the chiral limit is (Appendix A)
\begin{equation}
\theta_1=\frac{\ell+b}{2\pi}+O(a\Lambda\ell),\qquad \theta_2=-2C_1+O((a\Lambda)^2\ell),\qquad G^{-1}(0)=\frac1\pi+O(a\Lambda\ell).
\label{eq:tuning}
\end{equation}
$C_1=\bzint w/(\kappa^2+w^2)$ and $b$ are constants determined by the lattice kinetic term. The part $\ell/(2\pi)$ of $\theta_1$ is the part that sends the coupling to the critical point ($\theta_1=\infty$), and its coefficient $1/(2\pi)$ does not depend on the kinetic term. In natural coordinates, differences in the lattice kinetic term are limited to parallel shifts by the $\ell$-independent constants $b/(2\pi)$ and $-2C_1$. Moreover, since $\theta_2$ is independent of $\ell$ up to $O((a\Lambda)^2\ell)$, the coexistence line approaches the straight line $\theta_2=-2C_1$ in natural coordinates (Fig.~\ref{fig:overview}(a)). $b/(2\pi)$ is the ratio of $\Lambda$ parameters, and $-2C_1$ is the additive mass renormalization due to the Wilson term~\cite{AokiHigashijima1986}; both are known quantities, and Eq.~(\ref{eq:tuning}) shows where they appear in natural coordinates. The effect of the doublers (the three heavy fermions at the corners of the Brillouin zone) on long distances appears only through these two constants, $b$ and $C_1$.

From Eq.~(\ref{eq:tuning}), $\ell=2\pi\theta_1-b+O(a\Lambda\ell)$, so on the line of constant physics
\begin{equation}
\langle\sigma_n\rangle=a\Lambda=e^be^{-2\pi\theta_1}\big[1+O(a\Lambda\ell)\big].
\label{eq:transmutation}
\end{equation}
This is dimensional transmutation, $\Lambda=a^{-1}e^be^{-\pi/g^2}$, written in the mixed coordinates $(\theta_1,\langle\sigma_n\rangle)$, and dimensional transmutation becomes an exponential relation between the natural coordinate $\theta_1$ and the expectation coordinate $\langle\sigma_n\rangle$. When $a\to0$ is taken while the expectation coordinate is matched to $a\Lambda$ by the renormalization condition, the natural coordinate $\theta_1$ heads to the critical point $\theta_1=\infty$ at the rate $\log(1/(a\Lambda))/(2\pi)$. Differences in the kinetic term appear only in the prefactor $e^b$ of the exponential.

\section{How the kinetic-term directions are lost}
\label{sec:modes}
In this section, we study how distinguishable the kinetic-term direction $v_\delta$ is in the Fisher metric of the long-distance modes and in the Fisher metric of the full $P(\sigma)$. Both metrics are determined by the form of the propagator of $\sigma$ at all momenta.

\subsection{Propagator at all momenta}
On the line of constant physics, for $\hat q\gg a\Lambda$,
\begin{equation}
G^{-1}(q)\simeq\frac1\pi\big[\ell+\zeta(q)\big],\qquad \zeta(q)=\log\hat q+O(1)
\label{eq:Gall}
\end{equation}
($\zeta$ is a lattice function independent of $a\Lambda$; the error estimates are given in Appendix B). In physical units, $G\approx\pi/\log(k/\Lambda)$ for $\Lambda\ll k\ll1/a$, and $G\approx\pi/\ell$ at lattice-scale momenta. Short-wavelength fluctuations of $\sigma$ are weakened as $1/\ell$ by asymptotic freedom. Averaging over the Brillouin zone gives $1/\glat^2=(\ell+\gB)/\pi+O(a\Lambda\ell)$ ($\gB=\bzint\zeta=0.3715$). This $1/\ell$ is the source of the logarithms that appear later in the Fisher metric of the full $P(\sigma)$ and in the curvature.

\subsection{The kinetic-term directions degenerate in the Fisher metric of the long-distance modes}
\label{sec:phys}
For a kinetic-term direction to degenerate in the Fisher metric of the long-distance modes, two conditions are both necessary. On the line of constant physics, we split the component of the kinetic-term direction as $v_\delta G^{-1}(q)=v_\delta G^{-1}(0)+\Pi_\delta(q)$. Here $\Pi_\delta(q)=\partial_\tau[\Pi(q)-\Pi(0)]$ is the momentum-dependent part of the self-energy when the kinetic term is changed with the saddle-point value fixed. The first condition is the fixing of long-distance quantities, by which the momentum-independent part $v_\delta G^{-1}(0)$ is removed up to $O(a\Lambda\ell)$. If long-distance quantities are not fixed (in the limit with the bare couplings fixed), this part remains at $O(1)$. The second condition is the locality of the kinetic term, by which the momentum-dependent part $\Pi_\delta(q)$ starts at $q^2$. If the kinetic term is not local (for example, for long-range hopping that decays as a power), $\Pi_\delta(q)$ starts more slowly than $q^2$. Combining the two, at physical scales,
\begin{equation}
v_\delta G^{-1}(q)\simeq v_\delta G^{-1}(0)+c_\delta\hat q^2,\qquad v_\delta G^{-1}(0)=O(a\Lambda\ell)
\label{eq:vdecomp}
\end{equation}
(for deformations of $w$ that start at $p^4$ and for deformations of $\kappa$, $v_\delta G^{-1}(0)=O(a\Lambda)$; for deformations of $w$ that start at $p^2$, a term of order $\hat q^2\log(1/\hat q)$ is added to $c_\delta\hat q^2$). With the momentum $k$ fixed in physical units,
\begin{equation}
r_\delta(k)\sim
\begin{cases}
(a\Lambda)^2\log^2(1/(a\Lambda)) & (\text{deformations of}\ w\ \text{starting at}\ p^2,\ \text{d1}),\\
(a\Lambda)^2 & (\text{deformations of}\ w\ \text{starting at}\ p^4\ \text{and of}\ \kappa,\ \text{d2, d3})
\end{cases}
\label{eq:rphys}
\end{equation}
goes to zero. The estimates hold uniformly in the deformation (Appendix B).

For lattice readers, this result can be restated in the language of Symanzik's effective theory~\cite{Symanzik1983}. The amplitude $v_\delta G^{-1}=O(a\Lambda\ell)$ corresponds to the $O(a)$ errors of Wilson fermions. The rate at which it vanishes does not depend on the classical dimension of the operator of the deformation, because what dominates at physical scales is the momentum-independent part $v_\delta G^{-1}(0)$. A deformation of the kinetic term changes the additive mass renormalization ($C_1$) under the Wilson term and mixes with the mass term with a power-divergent coefficient. What remains after this part is absorbed by the tuning is $O(a)$. A deformation of $\kappa$ (d3) does not itself break chiral symmetry, but since it changes $C_1=\bzint w/(\kappa^2+w^2)$, it still mixes with the mass term, and $v_\delta G^{-1}(0)$ decreases in proportion to $a\Lambda$ (more slowly than $(a\Lambda)^2$; Appendix B).

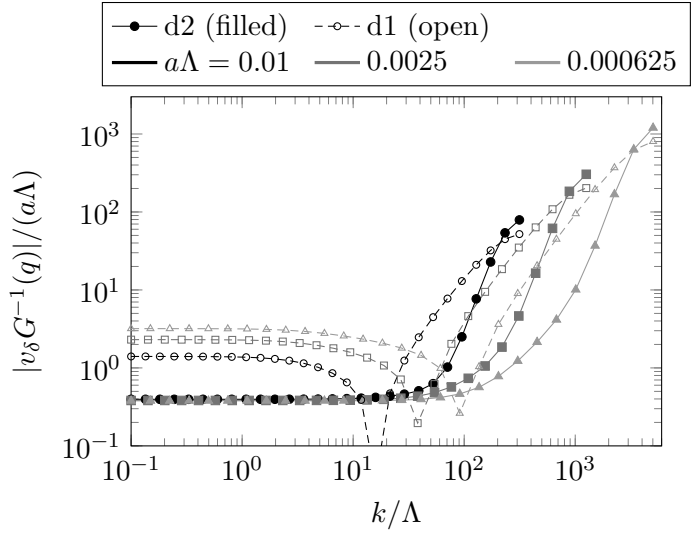
\begin{figure}[t]
\centering
\begin{tikzpicture}
\begin{axis}[xmode=log, ymode=log, xlabel={$k/\Lambda$}, ylabel={$|v_\delta G^{-1}(q)|/(a\Lambda)$},
  xmin=0.1, xmax=6000, ymin=0.1, ymax=3000, legend columns=3, legend style={at={(0.5,1.03)}, anchor=south, /tikz/every even column/.append style={column sep=0.6em}}, legend cell align=left]
\addplot[black, mark=*, mark size=1.6pt, forget plot] table[x=x,y=d2n]{data/v2_axis_1.dat};
\addplot[black!55, mark=square*, mark size=1.6pt, forget plot] table[x=x,y=d2n]{data/v2_axis_2.dat};
\addplot[black!40, mark=triangle*, mark size=1.8pt, forget plot] table[x=x,y=d2n]{data/v2_axis_3.dat};
\addplot[black, thin, mark=o, mark size=1.3pt, mark options={solid}, densely dashed, forget plot] table[x=x,y=d1n]{data/v2_axis_1.dat};
\addplot[black!55, thin, mark=square, mark size=1.3pt, mark options={solid}, densely dashed, forget plot] table[x=x,y=d1n]{data/v2_axis_2.dat};
\addplot[black!40, thin, mark=triangle, mark size=1.5pt, mark options={solid}, densely dashed, forget plot] table[x=x,y=d1n]{data/v2_axis_3.dat};
\addlegendimage{black, mark=*, mark size=1.6pt} \addlegendentry{d2 (filled)}
\addlegendimage{black, thin, mark=o, mark size=1.3pt, mark options={solid}, densely dashed} \addlegendentry{d1 (open)}
\addlegendimage{empty legend} \addlegendentry{}
\addlegendimage{black, line width=1.2pt} \addlegendentry{$a\Lambda=0.01$}
\addlegendimage{black!55, line width=1.2pt} \addlegendentry{$0.0025$}
\addlegendimage{black!40, line width=1.2pt} \addlegendentry{$0.000625$}
\end{axis}
\end{tikzpicture}
\caption{Momentum dependence of the component $v_\delta G^{-1}(q)$ of the kinetic-term direction on the line of constant physics. The vertical axis is the component divided by $a\Lambda$ (with the momentum along an axis), and the horizontal axis is the momentum in physical units, $k/\Lambda$. The plateaus of d2 (filled symbols) for the three values of $a\Lambda$ overlap at $0.38$--$0.40$, so $v_\delta G^{-1}(0)\propto a\Lambda$. The plateaus of d1 (open symbols) increase with $\ell$ as $1.40$, $2.30$, $3.19$, so $v_\delta G^{-1}(0)$ is of order $a\Lambda\ell$. The rise at the right end is due to $c_\delta\hat q^2$ and the lattice-scale part ($q\sim1$), and the value of $k/\Lambda$ at which $q$ approaches 1 differs for each $a\Lambda$. The downward dips of d1 are points where $v_\delta G^{-1}$ changes sign, and they appear because the absolute value is plotted on a logarithmic scale. The numbers were obtained by Gauss quadrature over the Brillouin zone (with a fine grid around $p=0$) and central differences with respect to the kinetic term.}
\label{fig:regimes}
\end{figure}

For deformations that start at $p^4$, the momentum-dependent part exceeds the momentum-independent part for $k\gtrsim k_*\sim[\Lambda/(a|c_\delta|)]^{1/2}$. $k_*$ is the boundary between the physical scale and the lattice scale; it goes to infinity in physical units and to zero in lattice units. The component of the kinetic-term direction thus has three regions. For $k\lesssim k_*$ the momentum-independent part dominates, for $k_*\lesssim k\ll1/a$ the term $c_\delta\hat q^2$ dominates, and for $k\sim1/a$ (the lattice scale) it takes an $a$-independent value of $O(1)$ (Fig.~\ref{fig:regimes}).

\subsection{\texorpdfstring{The kinetic-term directions remain in the Fisher metric of the full $P(\sigma)$}{The kinetic-term directions remain in the Fisher metric of the full P(sigma)}}
\label{sec:all}
At lattice-scale momenta, $G\approx\pi/\ell$ holds uniformly. Hence, in the Fisher metric of the full $P(\sigma)$, both the squared length of the mass-term direction and that of a kinetic-term direction become small as the same $1/\ell^2$. In their ratio $\ell$ cancels, and an $a$-independent value remains (Appendix B):
\begin{equation}
F_{mm}\simeq\frac{\pi^2}{2\ell^2},\qquad \frac{F_{\delta\delta}}{F_{mm}}\to\bzint(\Pi^{\rm lat}_\delta)^2,\qquad\frac{F^\perp_\delta}{F_{mm}}\to\VarB(\Pi^{\rm lat}_\delta).
\label{eq:global}
\end{equation}
$\Pi^{\rm lat}_\delta$ is the $a\Lambda\to0$ limit of $\Pi_\delta$, that is, the response of the self-energy of massless lattice fermions to the deformation of the kinetic term (hereafter the label lat denotes a quantity obtained by taking $a\to0$ with the momentum $q$ in lattice units fixed, that is, a quantity of massless lattice fermions). The values of the right-hand sides are $\bzint(\Pi^{\rm lat}_\delta)^2=0.264$ and $\VarB(\Pi^{\rm lat}_\delta)=0.0387$ for d1, $0.665$ and $0.190$ for d2, and $0.0050$ and $0.0036$ for d3 (obtained by fast Fourier transforms in position space; the digits shown do not change when the number of momentum points is varied from $1024^2$ to $8192^2$). At finite $a\Lambda$, $F_{\delta\delta}/F_{mm}$ is within 8\% of its limiting value for $\ell=3$--$9$ (Fig.~\ref{fig:modes}). Since the squared length of $v_\Lambda$ remains finite through the mean part $NG^{-1}(0)\to N/\pi$, compared with $v_\Lambda$ the kinetic-term directions also vanish, at the rate $1/(N\ell^2)$. The kinetic-term directions remain with the same magnitude as the mass-term direction when they are compared within the covariance part (the second term).

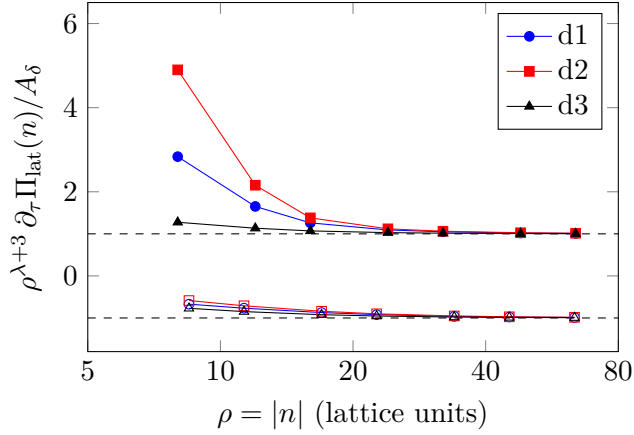
\begin{figure}[t]
\centering
\begin{tikzpicture}
\begin{axis}[xmode=log, xlabel={$\rho=|n|$ (lattice units)}, ylabel={$\rho^{\lambda+3}\,\partial_\tau\Pi_{\rm lat}(n)/A_\delta$},
  xmin=5, xmax=80, ymin=-1.8, ymax=6.5, xtick={5,10,20,40,80}, xticklabels={5,10,20,40,80}, legend pos=north east, legend cell align=left]
\addplot[blue, mark=*, mark size=1.8pt] table[x=rho, y expr=\thisrow{val}/1.21585]{data/tail_d1_axis.dat}; \addlegendentry{d1}
\addplot[red, mark=square*, mark size=1.8pt] table[x=rho, y expr=\thisrow{val}/(-2.43171)]{data/tail_d2_axis.dat}; \addlegendentry{d2}
\addplot[black, mark=triangle*, mark size=2pt] table[x=rho, y expr=\thisrow{val}/0.303964]{data/tail_d3_axis.dat}; \addlegendentry{d3}
\addplot[blue, mark=o, mark size=1.8pt, forget plot] table[x=rho, y expr=\thisrow{val}/1.21585]{data/tail_d1_diag.dat};
\addplot[red, mark=square, mark size=1.8pt, forget plot] table[x=rho, y expr=\thisrow{val}/(-2.43171)]{data/tail_d2_diag.dat};
\addplot[black, mark=triangle, mark size=2pt, forget plot] table[x=rho, y expr=\thisrow{val}/0.303964]{data/tail_d3_diag.dat};
\addplot[black, dashed, domain=5:80, forget plot] {1};
\addplot[black, dashed, domain=5:80, forget plot] {-1};
\end{axis}
\end{tikzpicture}
\caption{Far behavior of the response of the kinetic-term direction in position space, divided by the predicted amplitude $A_\delta$ of Eq.~(\ref{eq:tail}). Filled symbols are on an axis ($\varphi=0$) and open symbols are on the diagonal ($\varphi=\pi/4$), for which the predictions are $+1$ and $-1$ (dashed lines), respectively. Obtained by fast Fourier transforms with $4096^2$ points. The fractions of $\VarB(\Pi^{\rm lat}_\delta)$ from $|n|\le1$ are 0.976, 0.985, 0.510 for d1, d2, d3, and those from $|n|\le2$ are 0.999, 0.991, 0.959. The ratio of the values at $(3,4)$ and $(5,0)$, which have the same $|n|$, would be $-0.843$ for $\cos4\varphi$, but it is $-0.163$ for d1, and it approaches $-0.747$ for $(15,20)$ and $(25,0)$ and $-0.819$ for $(30,40)$ and $(50,0)$.}
\label{fig:tail}
\end{figure}

That $\Pi^{\rm lat}_\delta$ is not a constant in $q$ can be seen in position space. We write the propagator of massless fermions as $S_F=\Ds^{-1}$. The position-space form of the massless self-energy is $\Pi_{\rm lat}(n)={\rm tr}[S_F(n)S_F(-n)]$, and $\Pi^{\rm lat}_\delta$ transformed to position space becomes $\partial_\tau\Pi_{\rm lat}(n)$. By Parseval's identity, $\VarB(\Pi^{\rm lat}_\delta)=\sum_{n\neq0}|\partial_\tau\Pi_{\rm lat}(n)|^2$. Most of it comes from the change of correlations at a distance of one lattice spacing (nearest-neighbor lattice points). For example, for d1 the fraction from $|n|\le1$ is 0.976 (for the other directions, see the caption of Fig.~\ref{fig:tail}). The far tail is determined by the first term of the nonanalytic part, around $p=0$, of the change of the propagator due to the deformation of the kinetic term (Appendix C), and with $\rho=|n|$ and $\varphi$ the angle of $n$,
\begin{equation}
\partial_\tau\Pi_{\rm lat}(n)\simeq A_\delta\frac{\cos4\varphi}{\rho^{\lambda+3}},\qquad
A_\delta=\frac{12}{\pi^2}\ ({\rm d1},\ \lambda=3),\quad -\frac{24}{\pi^2}\ ({\rm d2},\ \lambda=3),\quad \frac3{\pi^2}\ ({\rm d3},\ \lambda=1)
\label{eq:tail}
\end{equation}
(Fig.~\ref{fig:tail}). This $\cos4\varphi$ form appears beyond 20--30 lattice spacings. The ratio of the tail to the main isotropic part of $\Pi_{\rm lat}(n)$ itself ($-1/(2\pi^2\rho^2)$) is $\rho^{-4}$ or $\rho^{-2}$, so at fixed physical distance $\rho a$ it goes to zero in proportion to $a^4$ or $a^2$. The far decay is determined by the structure of the anisotropy: d1 of dimension 3 and d2 of dimension 5 both decay as $1/\rho^6$, and d3 of dimension 4 decays as $1/\rho^4$.

\subsection{Contact-term direction}
\label{sec:contact}
The ratio in Sec.~IV C remains finite for general local deformations, but there is one concrete kinetic-term direction whose length goes to zero as $a\to0$ even in the Fisher metric of the full $P(\sigma)$. Since $\Ds^2=(w^2-\kappa^2)+\iu\gamma\cdot(2w\kappa)$,
\begin{equation}
\delta\Ds\propto\Ds^2,\qquad\text{i.e.,}\qquad\delta w=w^2-\kappa^2,\quad\delta\kappa_\mu=2w\kappa_\mu
\label{eq:contact}
\end{equation}
is a local, cubic-symmetric deformation in which $\delta w$ starts at $p^2$ and $\delta\kappa$ starts at $p^3$. The change of the massless propagator $S_F=\Ds^{-1}$ is $\delta S_F=-S_F\,\delta\Ds\,S_F\propto-1$, independent of momentum. In position space,
\begin{equation}
\delta S_F(n)\propto-\delta_{n,0},
\label{eq:contactn}
\end{equation}
and only the value of the propagator at coincident points, that is, the contact term, changes. Since $\Pi_{\rm lat}(n)={\rm tr}[S_F(n)S_F(-n)]$ also changes only at $n=0$, $\delta\Pi_{\rm lat}(n)\propto\delta_{n,0}$ is a constant in momentum space, and $\Pi^{\rm lat}_\delta\equiv0$. Hence the length of this direction goes to zero as $a\to0$ even in the Fisher metric of the full $P(\sigma)$. Seen from the two-point function of $\sigma$, as $a\to0$ this deformation only adds a contact term (a momentum-independent constant) to the self-energy, and it becomes indistinguishable from a parallel shift of $1/g^2$ (that is, of $\theta_1$). For families of conformal field theories, Kutasov showed that a change of contact terms in two-point functions is a change of coordinates of the couplings~\cite{Kutasov1989}. The direction (\ref{eq:contact}) is its lattice counterpart.

At finite $a\Lambda$, this deformation is not the same as a change of $1/g^2$. With $\delta\Ds=c\Ds^2$ ($c$ a constant), $\det(\Ds+c\Ds^2+\sigma)=\det(1+c\Ds)\det(\Ds+(1+c\Ds)^{-1}\sigma)$, so that, apart from a $\sigma$-independent factor, the coupling between the fermions and $\sigma$ changes from $\bar\psi\sigma\psi$ to $\bar\psi(1+c\Ds)^{-1}\sigma\psi$. At large $N$, this difference appears as an $O(\sigma_c)$ momentum-dependent part of $\delta S_F$, and the length is not exactly zero. On the line of constant physics, the ratio $F_{\delta\delta}/F_{mm}$ is $6.8\times10^{-5}$, $5.1\times10^{-6}$, $3.5\times10^{-7}$ at $a\Lambda=0.01$, $0.0025$, $0.000625$, going to zero roughly as $(a\Lambda)^2$ (at the same values of $a\Lambda$, d1 gives 0.256, 0.253, 0.253). We examined whether there are other directions that are degenerate in the Fisher metric of the full $P(\sigma)$, in the space spanned by eight local deformations including (\ref{eq:contact}) (the list of deformations is given in Appendix C). Of the eigenvalues of $\CovB(\Pi^{\rm lat}_\delta,\Pi^{\rm lat}_{\delta'})$ ($\delta$ and $\delta'$ run over the eight deformations), only one is zero, and its eigenvector coincides with (\ref{eq:contact}). Within the range examined, the only kinetic-term direction that is degenerate in the Fisher metric of the full $P(\sigma)$ is the contact-term direction. From the far tail (\ref{eq:tail}), one can show that $\Pi^{\rm lat}_\delta$ is not a constant, except for directions in which all three coefficients of the tail vanish (Appendix C). The direction (\ref{eq:contact}) is an example of such an excluded direction.

\subsection{Limit of the long-distance modes and effective dimension}
\label{sec:closed}
\begin{figure}[t]
\centering
\begin{tikzpicture}
\begin{axis}[ymode=log, xlabel={$\ell=\log(1/(a\Lambda))$}, ylabel={$F_{\delta\delta}/F_{mm}$, $r_\delta(k=\Lambda)$},
  xmin=3, xmax=9, ymin=1e-9, ymax=30, clip=false, legend columns=2, legend style={at={(0.5,1.03)}, anchor=south, /tikz/every even column/.append style={column sep=0.8em}}, legend cell align=left]
\addplot[blue, mark=*, mark size=1.8pt] table[x=L,y=rall1]{data/ratio_modes.dat}; \addlegendentry{full $P(\sigma)$, d1}
\addplot[red, mark=square*, mark size=1.8pt] table[x=L,y=rall2]{data/ratio_modes.dat}; \addlegendentry{full $P(\sigma)$, d2}
\addplot[blue, mark=o, mark size=1.8pt, mark options={solid}, dashed] table[x=L,y=rphys1]{data/ratio_modes.dat}; \addlegendentry{long-distance modes, d1}
\addplot[red, mark=square, mark size=1.8pt, mark options={solid}, dashed] table[x=L,y=rphys2]{data/ratio_modes.dat}; \addlegendentry{long-distance modes, d2}
\addplot[black, densely dotted, domain=3:9, samples=40, forget plot] {0.2113*(exp(-x)*x)^2};
\addplot[black, densely dotted, domain=3:9, samples=40, forget plot] {0.15645*exp(-2*x)};
\addplot[black, thin, domain=3:9, forget plot] {0.264};
\addplot[black, thin, domain=3:9, forget plot] {0.665};
\node[anchor=south west, font=\scriptsize, inner sep=1pt] at (axis cs:9,0.665) {0.665};
\node[anchor=north west, font=\scriptsize, inner sep=1pt] at (axis cs:9,0.264) {0.264};
\end{axis}
\end{tikzpicture}
\caption{How the kinetic-term directions appear in the two metrics. Symbols with solid lines are $F_{\delta\delta}/F_{mm}$ in the Fisher metric of the full $P(\sigma)$, and symbols with dashed lines are $r_\delta$ for the long-distance mode $k=\Lambda$. The numbers at the right edge (thin black lines) are the limiting values in the Fisher metric of the full $P(\sigma)$, and the black dotted lines are the predicted slopes for the long-distance modes, $(a\Lambda)^2\ell^2$ (d1) and $(a\Lambda)^2$ (d2), each matched at the point with the largest $\ell$.}
\label{fig:modes}
\end{figure}
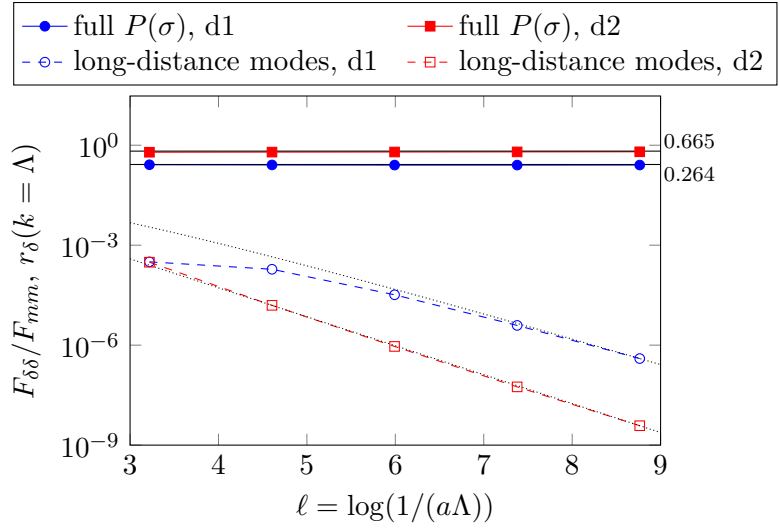

We examine what remains when all kinetic-term directions degenerate in the Fisher metric of the long-distance modes. Fixing the momentum $k$ in physical units, setting $q=ak$, and taking $a\Lambda\to0$, the components of each direction approach the following closed forms (Appendix E):
\begin{equation}
\begin{gathered}
G(q)\to G_c(k/\Lambda)=\frac{\pi\tanh u}u,\qquad a\Lambda\,v_\Lambda G^{-1}(q)\to\chi(k/\Lambda)=\frac1\pi\frac{2u}{\sinh2u},\\
v_mG^{-1}=1,\qquad v_\delta G^{-1}(q)\to0.
\end{gathered}
\label{eq:closed}
\end{equation}
Here $u={\rm arcsinh}(k/(2\Lambda))$, and the errors are $O(a\Lambda\ell)$. The metric $\frac12(d\log G)^2$ of a single mode has rank 1. Since $\chi$ is strictly monotonically decreasing, combining the limiting metrics of two or more modes with different $|k|$ gives rank 2, and the directions with nonzero length are spanned by $v_\Lambda$ and $v_m$. All kinetic-term directions have zero length in this limiting metric. That is, in the Fisher metric of the long-distance modes, the infinitely many kinetic-term directions all degenerate, and the only directions that do not degenerate are the two directions of the parameters $(\Lambda,m)$ of the family of continuum Gross--Neveu models. The statement in the Introduction, ``a finite-dimensional manifold remains for a renormalizable theory,'' holds in this sense. On the other hand, in the Fisher metric of the full $P(\sigma)$, the kinetic-term directions do not degenerate, except for the contact-term direction. Figure~\ref{fig:modes} shows the difference between the two metrics.

\section{Curvature and phase transitions}
\label{sec:curv}
\subsection{Curvature of the coupling plane}
The components of the metric depend on the choice of coordinates, but the curvature is a coordinate-independent quantity. In thermodynamic geometry, the magnitude of the scalar curvature of the thermodynamic metric is taken to be proportional to the correlation volume $\xi^d$, and it has been used as a coordinate-independent measure of the strength of interactions and of the closeness to a critical point~\cite{Ruppeiner1995}. Here we study how the curvature of the coupling plane behaves around the critical point approached in the continuum limit. An exponential family has two mutually dual connections, the e-connection, which is flat in the natural coordinates, and the m-connection, which is flat in the expectation coordinates~\cite{AmariNagaoka} (the names e and m come from the fact that the respective geodesics are linear combinations of $\log P$ (exponential) and of $P$ (mixture)). The $(\theta_1,\theta_2)$ plane at fixed kinetic term is a dually flat space in which the curvature of both connections vanishes, and what remains as curvature is the curvature of the Levi-Civita connection of the Fisher metric. In a dually flat space, the Levi-Civita connection in natural coordinates can be written as $\Gamma_{ij,k}=T_{ijk}/2$. Here $T_{ijk}=N_s^{-1}\partial_i\partial_j\partial_k\log Z$ is the Amari--Chentsov tensor (the third cumulant of the sufficient statistics), and the metric is $g_{ij}=N_s^{-1}\partial_i\partial_j\log Z$ (both per site). Hence the curvature is determined by an expression quadratic in $T$ with the inverse of $g$, and it expresses how large the third cumulant of the sufficient statistics is compared with the metric~\cite{AmariNagaoka}.

Since only the $\sigma_c$ direction of the metric (\ref{eq:metric}) is proportional to $N$, as $N\to\infty$ the sectional curvature approaches that of a warped product~\cite{ONeill}. Here the warped product is the metric obtained by adding, to the metric of the submanifold with $\sigma_c$ fixed, the metric $G^{-1}(0)\,d\sigma_c^2$ in the $\sigma_c$ direction (with the weight $G^{-1}(0)$ at each point) (Appendix D). The Gauss curvature of the $(\theta_1,\theta_2)$ plane at fixed kinetic term is $K=K_\infty+O(1/N)$, with
\begin{equation}
K_\infty=\frac{G(0)\big[G(0)\bzint G^2-2\bzint G^3\big]}{2\big(\bzint G^2\big)^2},
\label{eq:K}
\end{equation}
and for small $a\Lambda$ it becomes
\begin{equation}
K_\infty=\frac{G(0)^2}{2\glat^4}-\frac{G(0)}{\glat^2}-\frac32G(0)^2\VarB(\Pi)+O(1/\ell)=\frac12\ell^2+(\gB-1)\ell+c_K+O(1/\ell).
\label{eq:Kasym}
\end{equation}
\begin{sloppypar}Here $1/\glat^2=(\ell+\gB)/\pi+O(a\Lambda\ell)$, $\gB=0.3715$, and $c_K=\gB^2/2-\gB-(3\pi^2/2)\VarB(\Pi_{\rm lat})=-0.476$ ($\Pi_{\rm lat}$ is the massless self-energy, with $\VarB(\Pi_{\rm lat})=0.01169$). The $N^0$ part of the Fisher distance $s$ along the line of constant physics, measured from a point at finite $\ell$, diverges as $s=(1/\sqrt2)\log\ell+O(1)$, while the part proportional to $N$ is finite. Since the metric in the $\theta_1$ direction is bounded below by $\approx d\theta_1^2/(2\theta_1^2)$ independently of $\sigma_c$, the distance from a point at finite $\theta_1$ diverges along any curve heading to $\theta_1\to\infty$.\end{sloppypar}

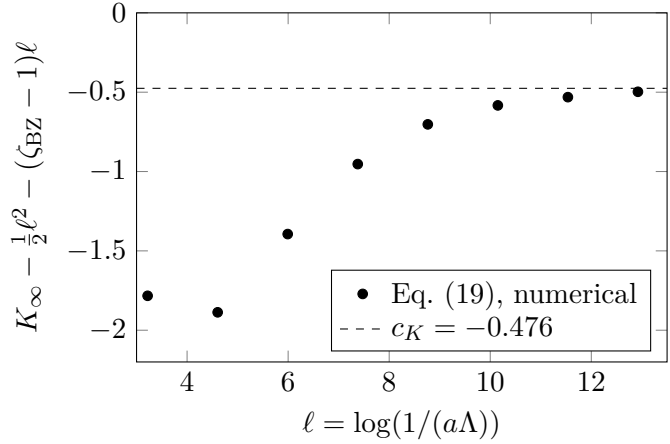
\begin{figure}[t]
\centering
\begin{tikzpicture}
\begin{axis}[xlabel={$\ell=\log(1/(a\Lambda))$}, ylabel={$K_\infty-\frac12\ell^2-(\gB-1)\ell$},
  xmin=3, xmax=13.5, ymin=-2.2, ymax=0, legend pos=south east, legend cell align=left]
\addplot[black, only marks, mark=*, mark size=1.8pt] table[x=L,y=R]{data/K_resid.dat}; \addlegendentry{Eq.~(\ref{eq:K}), numerical}
\addplot[black, dashed, domain=3:13.5] {-0.4756}; \addlegendentry{$c_K=-0.476$}
\end{axis}
\end{tikzpicture}
\caption{Remainder obtained by subtracting the first two terms of Eq.~(\ref{eq:Kasym}) from the curvature of the coupling plane ($\gB=0.3715$). The remainder is bounded and approaches the constant $c_K$ at large $\ell$. The deviation for $\ell\lesssim6$ is due to the $O(a\Lambda\ell^3)$ and $O(1/\ell)$ corrections (Appendix D).}
\label{fig:K}
\end{figure}

That is, the curvature diverges as $\frac12\log^2(1/(a\Lambda))$, and the critical point is infinitely far away from any point at finite $a$ when measured with the Fisher distance (Fig.~\ref{fig:K}). In coordinate-independent terms, $K_\infty\propto e^{2\sqrt2s}$ (the factor depends on the starting point of $s$), and the curvature grows exponentially with the Fisher distance traveled toward the critical point. The mechanism is as follows. When $\theta_1$ is moved with $\sigma_c$ fixed, $G^{-1}(0)=2\theta_1+\Pi(0)$ moves, and the weight $G^{-1}(0)^{1/2}$ of the warped product changes. On the other hand, the metric in the $\theta_1$ direction, $2\bzint G^2\,d\theta_1^2\approx d\ell^2/(2\ell^2)$, is small, so the weight changes quickly in Fisher distance and the curvature becomes large.

The same mechanism can also be written in the language of the dual structure. Whereas the Fisher metric is the connected two-point function of the operators that multiply the couplings, the Amari--Chentsov tensor is their connected three-point function. As $N\to\infty$, the components of the Amari--Chentsov tensor that remain in Eq.~(\ref{eq:K}) are only two: $T(v_\Lambda,v_\Lambda,v_m)=-N$, which comes from the saddle point, and $T(v_m,v_m,v_m)=-\bzint G^3$, which comes from the one-loop fluctuations. Using the components normalized by the lengths of the directions, $|v_\Lambda|^2=NG^{-1}(0)$ and $|v_m|^2=F_{mm}=\frac12\bzint G^2$, as $\hat T_{abc}=T(v_a,v_b,v_c)/(|v_a||v_b||v_c|)$, we obtain
\begin{equation}
K_\infty=\frac14\Big(\hat T_{\Lambda\Lambda m}^2-\hat T_{\Lambda\Lambda m}\hat T_{mmm}\Big),
\label{eq:KT}
\end{equation}
\begin{sloppypar}\noindent where the first term gives $G(0)^2/(2\bzint G^2)$ in Eq.~(\ref{eq:K}) and the second term gives $-G(0)\bzint G^3/(\bzint G^2)^2$ (Appendix D). The divergence $\frac12\ell^2$ comes from the first term. The connected three-point function $T(v_\Lambda,v_\Lambda,v_m)$ is independent of $a$, but since the connected two-point function $F_{mm}\approx\pi^2/(2\ell^2)$ goes to zero, the normalized component $\hat T_{\Lambda\Lambda m}\approx-\sqrt2\,\ell$ diverges. $\hat T_{mmm}$ approaches $-2\sqrt2$ and stays finite.\end{sloppypar}

\subsection{Planes including kinetic-term directions}
In planes that include kinetic-term directions, the curvature grows with a different magnitude. The Fisher metric among kinetic-term directions approaches the covariance $\CovB(\Pi^{\rm lat}_\delta,\Pi^{\rm lat}_{\delta'})$ of the responses to the deformations of the kinetic term multiplied by $\glat^4/2\approx\pi^2/(2\ell^2)$. That is, the space of kinetic-term directions shrinks by the factor $\pi/(\sqrt2\ell)$ while keeping its shape, and its curvature grows as $\ell^2$ (Appendix D).

\subsection{Curvature and phase transitions}
The continuum limit requires a divergent correlation length, that is, a continuous transition. We describe how the divergence of the curvature is related to it in this model (derivations are given in Appendix D). In two dimensions, the scalar curvature is $R=2K$, and in Ruppeiner's convention the sign is reversed ($R=-2K$)~\cite{Ruppeiner1995}. Below we compare only powers of absolute values. The coexistence line is a line of first-order transitions, across which $\langle\sigma_n\rangle$ jumps by $2a\Lambda[1+O(a\Lambda\ell)]$. At each point on the line, the curvature computed from one phase is finite, and the metric is discontinuous across the line. The curvature diverges only at the endpoint of the line ($g^2\to0$), where the jump vanishes and the correlation length of $\sigma$ becomes $\xi/a=1/(2a\Lambda)\to\infty$. Hence, on the line of constant physics, the point at which the continuous transition occurs coincides with the point at which the curvature diverges. On the other hand, neither the part of the metric proportional to $N$ nor the $N^0$ part diverges. The part proportional to $N$ is finite with $G^{-1}(0)\to1/\pi$ (the susceptibility of $\sigma$ is $G(0)\to\pi$), and the curvature, $K_\infty\approx(\pi^2/2)/\glat^4$, is determined by the effective coupling $\glat^2$ at the lattice scale.

The argument that derives $|R|\sim\xi^d$ in thermodynamic geometry uses the fact that the singular part of the free energy dominates the metric~\cite{Ruppeiner1995}. If $\Delta$ is the dimension of the operators conjugate to the coordinates ($\sum_n\sigma_n^2$ and $\sum_n\sigma_n$, which multiply the natural coordinates $\theta_1$ and $\theta_2$), the singular part of the susceptibility has size $\xi^{d-2\Delta}$, so without an operator with $2\Delta<d$ the singular part does not dominate the metric. In this model, neither $\sigma$ ($\Delta=1=d/2$) nor the fluctuation of $\sigma^2$ ($\Delta=2$), which are conjugate to the coordinates, satisfies this condition. Indeed, the singular part is only a fraction $(a\Lambda)^2$ of the metric, and the curvature is determined by the lattice-scale modes (the regular part) and grows as $(\log\xi)^2$ because of asymptotic freedom. The mechanism itself, in which the singular part disappears and the regular part determines the curvature, was described by Janke \textit{et al.} for models with a negative specific-heat exponent~\cite{JJK2003} (equivalent to $2\Delta>d$ for the thermal operator). For the quantum geometric tensor as well, Campos Venuti and Zanardi showed that criticality alone does not guarantee a singular divergence of the metric and that the condition is determined by the dimensions of operators~\cite{CamposVenutiZanardi2007}. However, being an essential singularity is not the reason for $(\log\xi)^2$. The zero-temperature critical point of the one-dimensional Ising model is an essential singularity, but the susceptibility of the spin ($\Delta=0$) diverges and $|R|\sim\xi^d$ holds~\cite{DJK2002}. Also, the statement ``the curvature diverges if there is a phase transition'' does not hold in general, and there are counterexamples in both directions~\cite{Dolan1998,RSSS2012,BrodyHook2009,EGJ2020}.

\section{Changes of scheme and the characterization of field theories}
\label{sec:scheme}
To answer the last question of the Introduction (whether a change of scheme is a gauge transformation), we divide changes of scheme into two kinds. S1 is a change of the renormalization condition, and S2 is a change of the regularization, that is, of the kinetic term. As an example of S1, if the renormalization condition is replaced by the condition that fixes the value of the inverse propagator at momentum $k_R$, $G^{-1}(ak_R)=c_R$ (a momentum-subtraction condition), this only assigns to the same points on the coexistence line a $\Lambda$ determined as the solution of $G_c^{-1}(k_R/\Lambda)=c_R$ ($G_c$ is given in Eq.~(\ref{eq:closed})), with corrections of $O(a\Lambda\ell)$ (Appendix E). As candidates for ``the theory'' in ``does not change the theory,'' we consider three: (a) the statistical manifold itself, (b) the distribution at lattice spacing $a$, and (c) the limit of the distribution of the long-distance modes. We write the set of Fisher metrics of the long-distance modes (the limits for each momentum $k$) as $\{F^c_k\}$.

\begin{table}[t]
\centering
\begin{tabular}{lll}
\toprule
Candidate for ``the theory'' & \shortstack[l]{S1 (change of\\renormalization condition)} & \shortstack[l]{S2 (change of\\kinetic term)}\\
\midrule
(a) Statistical manifold itself & unchanged & unchanged\\
(b) Distribution at lattice spacing $a$ & \shortstack[l]{only the assignment\\of $\Lambda$ changes} & changes\\
(c) \shortstack[l]{Limit of the distribution\\of the long-distance modes} & unchanged & unchanged\\
\bottomrule
\end{tabular}
\caption{What changes under a change of scheme.}
\label{tab:scheme}
\end{table}

(a) does not change under either change, since only the choice of points on the same manifold changes. For (b), S1 only reassigns the value of $\Lambda$ to the same set of points, the coexistence line. The S2 directions all degenerate in the Fisher metrics of the long-distance modes $\{F^c_k\}$ (Secs.~IV B and IV E) and do not degenerate in the Fisher metric of the full $P(\sigma)$, except for the contact-term direction (\ref{eq:contact}) (Secs.~IV C and IV D). Since S1 only reassigns $\Lambda$ to the same distribution, it is a change of coordinates even at finite $a$. The contact-term direction becomes indistinguishable from a change of the coupling $1/g^2$ as $a\to0$ (Table~\ref{tab:scheme}). For changes of fields, the Fisher metric does not change under transformations by sufficient statistics~\cite{AmariNagaoka}. Moreover, $\{F^c_k\}$ determines $G_c$ up to a constant factor (the normalization of the field) (Appendix E). From the above, the Fisher metric of the full $P(\sigma)$ does not identify general S2 and therefore does not characterize the field theory, while $\{F^c_k\}$ identifies changes of fields, changes of coordinates, S1, and S2, and distinguishes different values of $\Lambda$ and $m$.

Hence the last question of the Introduction can be answered as follows. ``The theory'' in the phrase ``changing the scheme does not change the theory'' corresponds to the limit of the distribution of the long-distance modes. The distribution at lattice spacing $a$ changes when the kinetic term is changed. The view that a change of scheme is a gauge transformation holds for S1 on the whole statistical manifold even at finite $a$. For the contact-term direction, it holds in the $a\to0$ limit even in the Fisher metric of the full $P(\sigma)$, and for general S2 it holds only in the Fisher metric of the long-distance modes. At large $N$, the distribution of $\sigma$ is determined by $G_c$ and $\sigma_c$, so the characterization by $\{F^c_k\}$ has the same content as the usual definition of the continuum limit as the convergence of correlation functions in physical units.

\section{Discussion}
We describe the relation between our results and previous work. That the Fisher metric is a two-point function of operators and that irrelevant directions shrink under coarse graining are found in Refs.~\cite{Machta2013,Raju2018,BHM2015}. These works, however, coarse grain the observation with the lattice fixed and do not treat $a\to0$ along the line of constant physics. In this paper, we do not coarse grain, and we separated the contributions of the physical scale and the lattice scale by the exact mode decomposition at large $N$. The idea of regarding degenerate directions as irrelevant and the idea of regarding a change of regularization as a move within an equivalence class are found in Bény and Osborne~\cite{BenyOsborneNJP}. We showed this as a degeneracy in the continuum limit without using a coarse-graining channel and a threshold, and we further showed that in the Fisher metric of the full $P(\sigma)$ these directions do not degenerate in general. For families of conformal field theories, a change of contact terms is a change of coordinates, and the choice of renormalization conditions is a choice of coordinates~\cite{Kutasov1989}. Our S1 is an example of this, and the contact-term direction (\ref{eq:contact}) is a lattice version of Kutasov's result in the sense that a change of the contact term of the propagator becomes indistinguishable from a change of the coupling $1/g^2$ as $a\to0$. The other S2 (irrelevant directions) degenerate only in the Fisher metric of the long-distance modes and are outside Kutasov's framework. There is no contradiction with the statement of Pagani and Sonoda that ``there is no standard metric on theory space''~\cite{PaganiSonoda2018}; that the Fisher metric of the full $P(\sigma)$ depends on the scheme and that a scheme-independent metric is obtained in the limit of the long-distance modes are the two sides of it. On curvature and phase transitions, that the curvature is finite on a line of first-order transitions and diverges at the critical point has precedents in Refs.~\cite{Dolan1998,RSSS2012}, and the mechanism by which the regular part determines the curvature is found in Ref.~\cite{JJK2003}. What is new in this paper is that the endpoint is an asymptotically free ultraviolet fixed point, that the curvature grows as $(\log\xi)^2$ while the metric does not diverge, and that the curvature grows exponentially with the Fisher distance traveled toward the critical point. Stout discussed the mechanism by which points at infinity appear in the information metrics of families of field theories from the viewpoint of distinguishability, and stated that points at infinity do not appear near typical quantum critical points~\cite{Stout2021}. The critical point in this paper is an asymptotically free ultraviolet fixed point, where the Fisher distance diverges as $\log\log(1/(a\Lambda))$.

Our results are limited to the large-$N$ theory, two dimensions, and the neighborhood of Wilson fermions (with the Wilson-term coefficient 1). We take $N\to\infty$ first and have not examined whether the $1/N$ corrections are uniform in $\ell$. Since logarithmic factors appear in the $1/N$ corrections of the two-dimensional Gross--Neveu model, this point is not trivial. The Fisher metric depends on the choice of the observed random variable (data-processing inequality). We have not examined which of our conclusions change when, for example, $\bar\psi\psi$ is taken as the observable instead of the auxiliary field $\sigma$. That the only direction that is degenerate in the Fisher metric of the full $P(\sigma)$ is the contact-term direction has only been checked numerically in an eight-dimensional space. For the massive case, we examined the tuning formula only at the first order in $a$.

\section{Summary and future problems}
With this paper, the continuum limit of lattice field theory can now be discussed in the language of information geometry. The family of lattice actions is, at first sight, an infinite-dimensional statistical manifold. On it, the operation of removing the regularization while imposing renormalization conditions is represented as a curve heading to the critical point at infinity, and by measuring with the Fisher metric, one can follow how the infinitely many irrelevant directions become indistinguishable. In the large-$N$ limit of the two-dimensional lattice Gross--Neveu model with Wilson fermions, the mode decomposition of the Fisher metric allowed us to follow this process analytically. In the information obtained by measuring only long-distance correlations, that is, in the Fisher metric of the long-distance modes, only the two directions $(\Lambda,m)$ of the continuum theory remain, and the fact that a renormalizable theory is determined by a finite number of parameters appeared in the form that only a finite-dimensional part of the infinite-dimensional manifold can be distinguished. In the Fisher metric of the full $P(\sigma)$, differences in the kinetic term, except for the contact-term direction, remain in correlations at distances of the order of the lattice spacing, and fixing the modes and $a\to0$ do not commute. A change of the renormalization condition is a change of coordinates on the statistical manifold even at finite $a$, and the contact-term direction becomes indistinguishable from a change of the coupling as $a\to0$. The critical point is infinitely far away in Fisher distance from any point at finite $a$, and the curvature grows exponentially with the distance traveled toward it.

These results lead to the following five problems. The first is theories that are not asymptotically free. The three-dimensional Gross--Neveu model (more generally, $2<d<4$) has a nontrivial ultraviolet fixed point at large $N$ and is regarded as a simple example of asymptotic safety~\cite{BGS2011}. Since the ratios that remain in the Fisher metric of the full $P(\sigma)$ and the $\frac12\ell^2$ of the curvature are due to the $1/\ell$ (\ref{eq:Gall}) that comes from asymptotic freedom, their behavior should change near an ultraviolet fixed point. The second is the $1/N$ corrections. How far the results of this paper, obtained by taking $N\to\infty$ first, hold at finite $N$ depends on whether the $1/N$ corrections are uniform in $\ell$. The third is the relation to field redefinitions. Since the contact-term direction is a deformation close to a field redefinition, we will study how its correspondence to a change of coordinates changes when the Jacobian of the measure that accompanies the redefinition shifts the couplings (in the presence of anomalies; Fujikawa's method~\cite{Fujikawa1979}). The fourth is the relation to the problems of constructing continuum limits mentioned in the Introduction. The Fisher metric of the long-distance modes characterizes the family of limiting distributions as a finite-dimensional statistical manifold. For example, can the triviality of the four-dimensional $\phi^4$ theory~\cite{AizenmanDuminilCopin2021} be read as the vanishing of the length, in the Fisher metric of the long-distance modes, of the direction that moves the coupling with the renormalized mass fixed? The fifth is the relation to Amari's theory of curved exponential families. The distribution of the large-$N$ theory is a submanifold of the family of Gaussian distributions whose natural coordinates are the precisions $NG^{-1}(q)$ of the modes. Since $v_m$ shifts $G^{-1}$ by a constant, it is an e-geodesic of this larger family, and the kinetic-term directions bend the submanifold. We will study how the kinetic-term directions are lost in the language of the curvature and information loss of curved exponential families~\cite{Efron1975,Amari1982}.

\section*{Data availability}
The code and data for the numerical calculations will be made public after the publication of this paper.

\begin{acknowledgments}
This work was encouraged by the seminar given by Takato Mori at Tokyo Woman's Christian University on October 1, 2026~\cite{MoriNagasawa2026}, and by the subsequent discussions. The author thanks Akio Fujiwara for the lectures on information geometry that the author attended in 2012 at the Advanced Core Course of Mathematics and Physics at Osaka University, for the textbook based on them~\cite{Fujiwara2015}, and for the exchange in 2015 that encouraged the author to study renormalization from the viewpoint of information geometry. This work was supported by JSPS KAKENHI Grants No. 20K14479, No. 22H05111, No. 22K03539, No. 22H05112, and JST BOOST Grant No. JPMJBY24F1. This work was supported by JST, ARiSE, Japan Grant Number JPMJAR2633. Anthropic's Claude Opus 5.5 was used, under the author's direction, for the analytical derivations and numerical calculations, the literature survey, and the preparation of the manuscript. The author assumes responsibility for the scientific claims, numerical results, citations, and final content of the manuscript.
\end{acknowledgments}

\appendix
\section{Norms of deformations and the line of constant physics}
\label{app:lcp}
We measure the size of a deformation with the following norms:
\begin{equation}
|\delta w|\le\|\delta\|_1w_W,\qquad|\delta w|\le\|\delta\|_2w_W^2,\qquad|\delta\kappa_\mu|\le\|\delta\|_3|\hat p_\mu|^3.
\label{eq:norms}
\end{equation}
The first is for deformations of $w$ that start at $p^2$, as the Wilson term does, the second for deformations of $w$ that start at $p^4$, and the third for deformations of $\kappa$ that start at $p^3$; we write $\|\delta\|=\|\delta\|_1+\|\delta\|_2+\|\delta\|_3$. That an estimate in the main text holds ``uniformly in the deformation'' means that the estimate is proportional to $\|\delta\|$ with a proportionality constant independent of the deformation. In the following appendices, $C$ denotes a constant independent of $a\Lambda$, $q$, and the deformation, whose value may change from equation to equation.

We write $M=\sigmabar+w$, $D=\kappa^2+M^2$, $I(\sigmabar)=\bzint\log D$, and $J=I'=\bzint2M/D$. Since the Wilson term lifts the doublers, $D\ge\hat p^2+\sigmabar^2$ ($\hat p^2=\sum_\mu\hat p_\mu^2$) for $\sigmabar\ge0$, and $D$ vanishes only at $p=0$ and $\sigmabar=0$. In $J''=\bzint\mathcal J$, $\mathcal J=4M(M^2-3\kappa^2)/D^3$, the integrand of the continuum theory, $\mathcal J_c=4\sigmabar(\sigmabar^2-3p^2)/(p^2+\sigmabar^2)^3$, gives $\int\mathcal J_c\,d^2p/(2\pi)^2=-1/(\pi\sigmabar)$. The integral of the difference from the lattice is $O(\log(1/|\sigmabar|))$, because $D$ is bounded below in the ultraviolet and $\mathcal J$ is a homogeneous function of degree $-3$ in the infrared. Integrating, we obtain ($J_1$ is a constant determined by the kinetic term)
\begin{equation}
J'(\sigmabar)=\frac1\pi\log\frac1{|\sigmabar|}+J_1+O\Big(\sigmabar\log\frac1{|\sigmabar|}\Big).
\label{eq:Jp}
\end{equation}

The line of constant physics is determined by the two stationarity conditions $(\sigma_\pm-m_0)/g^2=J(\sigma_\pm)$ and the coexistence condition $V(\sigma_+)=V(\sigma_-)$ (the unknowns are $m_0$, $g^2$, and $\sigma_-$). Measuring $\sigmabar$ in units of $a\Lambda$ and keeping, in the expansion of $J$, only $J(0)$, the linear term in $\sigmabar$, and the term that comes from the logarithm in Eq.~(\ref{eq:Jp}), the integrand of the coexistence condition becomes an odd function of $\sigmabar/(a\Lambda)$, and $\sigma_-=-a\Lambda$ is a solution. The remaining terms shift this solution by a relative amount $O(a\Lambda\ell)$. Hence $\sigma_-=-a\Lambda[1+O(a\Lambda\ell)]$ and $G^{-1}(0)=V''(\sigma_+)=1/g^2-J'(a\Lambda)=1/\pi+O(a\Lambda\ell)$.

Derivation of Eq.~(\ref{eq:tuning}). From $G^{-1}(0)=1/\pi+O(a\Lambda\ell)$ and Eq.~(\ref{eq:Jp}), $1/g^2=J'(a\Lambda)+1/\pi+O(a\Lambda\ell)=\ell/\pi+J_1+1/\pi+O(a\Lambda\ell)$, so $\theta_1=(\ell+b)/(2\pi)+O(a\Lambda\ell)$ with $b=1+\pi J_1$. The stationarity and coexistence conditions give $m_0/g^2=-J(0)+O((a\Lambda)^2\ell)$, and $J(0)=\bzint2w/(\kappa^2+w^2)=2C_1$.

\section{Propagator and the responses of kinetic-term directions}
\label{app:resp}
We write the self-energy at the saddle-point value $\sigmabar$ as $\Pi(q;\sigmabar)$ ($\Pi_{\rm lat}(q)=\Pi(q;0)$), and with the integrand $f(p_1,p_2)=2(M_1M_2-\kappa_1\cdot\kappa_2)/(D_1D_2)$, we write $\Pi(q;\sigmabar)=\bzint f(p,p+q)$. Since $f=2(Y_1\cdot\Omega Y_2)/(|Y_1|^2|Y_2|^2)$ ($Y_i=(\kappa_i,M_i)$, $\Omega={\rm diag}(-1,-1,1)$), homogeneity gives, for derivatives of orders $j$ and $j'$,
\begin{equation}
|\partial^j_{Y_1}\partial^{j'}_{Y_2}f|\le C D_1^{-(1+j)/2}D_2^{-(1+j')/2}.
\label{eq:homog}
\end{equation}

Derivation of the propagator (\ref{eq:Gall}) and its errors. We use the identity $G^{-1}(q)=V''(\sigma_+)+\Pi(q;a\Lambda)-\Pi(0;a\Lambda)$. For $|q|\le Ca\Lambda$, splitting the integral of the symmetrized second difference into infrared and ultraviolet parts gives $\Pi(q;a\Lambda)-\Pi(0;a\Lambda)=\Delta\Pi_c(q)+O(a\Lambda)$. Here $\Delta\Pi_c(q)=\frac1{4\pi}[(q^2+4(a\Lambda)^2)\int_0^1dt/((a\Lambda)^2+t(1-t)q^2)-4]\ge0$ is the difference in the continuum theory. For $\hat q\ge a\Lambda$, Eq.~(\ref{eq:homog}) gives $|\partial_\sigmabar\Pi(q;\sigmabar)|\le(C/\hat q)(1+\log(\hat q/\sigmabar))$, so that $\Pi_{\rm lat}(q)$ exists and $|\Pi(q;a\Lambda)-\Pi_{\rm lat}(q)|\le C(a\Lambda/\hat q)(2+\log(\hat q/(a\Lambda)))$. Also, from Eq.~(\ref{eq:Jp}), $\Pi(0;a\Lambda)=-J'(a\Lambda)=-\ell/\pi-J_1+O(a\Lambda\ell)$. Inserting these into the identity, for $\hat q\ge a\Lambda$ we obtain
\begin{equation}
G^{-1}(q)=\frac1\pi\big[\ell+\zeta(q)\big]+O\Big(a\Lambda\ell+\frac{a\Lambda}{\hat q}\big(1+\log\frac{\hat q}{a\Lambda}\big)\Big),\qquad|\zeta(q)-\log\hat q|\le C.
\label{eq:Gbound}
\end{equation}
Here $\zeta(q)=1+\pi(\Pi_{\rm lat}(q)+J_1)$, and $\gB=\bzint\zeta=1+\pi(2C_1^2+J_1)$ ($\bzint\Pi_{\rm lat}=\Pi_{\rm lat}(n=0)=2C_1^2$). At all $q$, $G(q)$ is bounded above and below by constant multiples of $1/(1+\log(1+\hat q/(a\Lambda)))$.

Momentum-independent part. $v_\delta G^{-1}(0)=\partial_\tau V''(\sigma_+)$, and $\tau$ enters only through the remainder of the expansion in Appendix A. From $\partial_\tau J''=\bzint\big(\partial_M\mathcal J\,\delta w+\partial_\kappa\mathcal J\cdot\delta\kappa\big)$ and $|\partial\mathcal J|\le C/D^2$, $|\partial_\tau J''|\le C\|\delta\|_1\log(1/\sigmabar)$ for the type with $|\delta w|\le\|\delta\|_1D/2$, and $C(\|\delta\|_2+\|\delta\|_3)$ for the other types. Integrating this twice and inserting it into the shift of the solution in Appendix A, we obtain
\begin{equation}
|v_\delta G^{-1}(0)|\le C\big[\|\delta\|_1a\Lambda\ell+(\|\delta\|_2+\|\delta\|_3)a\Lambda\big].
\label{eq:vzero}
\end{equation}
Numerically, for $a\Lambda=0.04$ to $0.000625$, $v_\delta G^{-1}(0)/(a\Lambda)$ is $0.38$--$0.44$ for d2 and $-0.33$ to $-0.27$ for d3, proportional to $a\Lambda$, and for d1 it is of order $a\Lambda\ell$ (Fig.~\ref{fig:regimes}).

Derivation of Eq.~(\ref{eq:rphys}). We write $\Pi_\delta(q;\sigmabar)=\bzint\big(\mathcal A(p;q)\delta w(p)+\mathcal B_\mu(p;q)\delta\kappa_\mu(p)\big)$ ($\mathcal A=\partial_{M_1}[f(p,p+q)+f(p,p-q)]-2\partial_{M_1}f(p,p)$, etc.). Where $p$ is away from $q$, we expand $\mathcal A$ and $\mathcal B_\mu$ to second order in $q$, and we integrate separately around $p=0$ and $p=\mp q$; this gives $|\Pi_\delta|\le C(\|\delta\|_1+\|\delta\|_3)[\hat q^2(1+\log\frac1{\hat q})+\sigmabar(1+\log^+\frac{\hat q}\sigmabar)]$ for deformations of $w$ that start at $p^2$ and for deformations of $\kappa$, and $C\|\delta\|_2(\hat q^2+\sigmabar^2)$ for deformations that start at $p^4$. Integrating $\partial_\sigmabar\Pi_\delta$, we further obtain $|\Pi_\delta(q;a\Lambda)-\Pi^{\rm lat}_\delta(q)|\le C\|\delta\|a\Lambda(1+\ell)$. Together with Eq.~(\ref{eq:vzero}), at all $q$,
\begin{equation}
r_\delta(q)\le C\|\delta\|^2\Big[a\Lambda\ell+\hat q^2\big(1+\log\tfrac1{\hat q}\big)+a\Lambda\big(1+\log^+\tfrac{\hat q}{a\Lambda}\big)\Big]^2
\label{eq:bound}
\end{equation}
holds. For $k$ in physical units, $\hat q\sim a\Lambda$, so $r_\delta\le C\|\delta\|^2(a\Lambda)^2\ell^2$. For directions that do not change the coefficient of the Wilson term, $\ell^2$ is replaced by $(1+\log^+(k/\Lambda))^2$.

Derivation of Eq.~(\ref{eq:global}). From Eq.~(\ref{eq:Gbound}), $G=(\pi/\ell)[1+O((1+|\log\hat q|)/\ell)]$ holds uniformly for $\hat q\ge(a\Lambda)^{1/2}$, and in the region $\hat q<(a\Lambda)^{1/2}$, $G\le C$ on an area of $O(a\Lambda)$. Using $\bzint(1+|\log\hat q|)^2(\Pi^{\rm lat}_\delta)^2<\infty$ and $|v_\delta G^{-1}-\Pi^{\rm lat}_\delta|\le C\|\delta\|a\Lambda\ell$, Eq.~(\ref{eq:global}) is obtained with $O(1/\ell)$ corrections.

\section{Far behavior}
\label{app:tail}
Derivation of Eq.~(\ref{eq:tail}). By the convolution theorem, the $q$-dependent part of the massless self-energy $\Pi_{\rm lat}(q)=\bzint{\rm tr}[S_F(p)S_F(p+q)]$ is the Fourier transform of $\Pi_{\rm lat}(n)={\rm tr}[S_F(n)S_F(-n)]$ ($n\neq0$). Hence $\Pi^{\rm lat}_\delta$ being a constant is equivalent to $\partial_\tau\Pi_{\rm lat}(n)=0$ for all $n\neq0$. Splitting $S_F(n)=-\iu\gamma_\mu S_\mu(n)+S_W(n)$ ($S_\mu$ and $S_W$ are the Fourier transforms of $\kappa_\mu/D$ and $w/D$, with $D=\kappa^2+w^2$) gives $\Pi_{\rm lat}(n)=-2S_\mu(n)S_\mu(-n)+2S_W(n)S_W(-n)$.

For the Wilson kinetic term, $D(p)>0$ for $p\neq0$, so $S_F(p)$ and $\partial_\tau S_F(p)$ are analytic on the Brillouin zone except at $p=0$ and can be expanded around $p=0$ in a sum of homogeneous terms. The expansion as $|n|\to\infty$ of the Fourier coefficients of such a function is determined by the Fourier transforms of the nonanalytic terms in the expansion at $p=0$ (terms of homogeneous degree $\lambda$ and angular order $j>\lambda$ with $j-\lambda$ even), and the rest decays faster than any power (a two-dimensional version of Lighthill's method~\cite{Lighthill1958}). Writing the momentum in polar coordinates as $p=|p|(\cos\varphi_p,\sin\varphi_p)$, the Fourier transform of a homogeneous term is
\begin{equation*}
\int\frac{d^2p}{(2\pi)^2}e^{\iu p\cdot n}|p|^\lambda e^{\iu j\varphi_p}=C_{\lambda,j}\frac{e^{\iu j\varphi}}{\rho^{\lambda+2}},\qquad C_{\lambda,j}=\frac{\iu^j}{2\pi}2^{\lambda+1}\frac{\Gamma(\frac{j+\lambda+2}2)}{\Gamma(\frac{j-\lambda}2)}.
\end{equation*}
We write the vector part of the propagator as a complex number, $S_c=S_{\mu=1}+\iu S_{\mu=2}$ ($(\kappa_1+\iu\kappa_2)/D$ in momentum space). The leading term of $S_c$ is $e^{\iu\varphi_p}/|p|$, and $S_c(n)=\frac{\iu}{2\pi}e^{\iu\varphi}/\rho+\cdots$. If the leading nonanalytic term of the change of $S_c$ due to the deformation of the kinetic term is $\alpha|p|^\lambda e^{5\iu\varphi_p}$, then $\delta S_c(n)=\alpha C_{\lambda,5}e^{5\iu\varphi}/\rho^{\lambda+2}+\cdots$, and from $\partial_\tau\Pi_{\rm lat}=-4{\rm Re}[S_c\overline{\delta S_c}]+\cdots$ we obtain
\begin{equation*}
\partial_\tau\Pi_{\rm lat}(n)=-\frac{\alpha\,2^{\lambda+1}}{\pi^2}\frac{\Gamma(\frac{\lambda+7}2)}{\Gamma(\frac{5-\lambda}2)}\frac{\cos4\varphi}{\rho^{\lambda+3}}+O(\rho^{-\lambda-5}).
\end{equation*}
The nonanalytic part of $S_W$ starts at degree 2 ($S_W(n)\sim\rho^{-4}$), and $S_W(n)S_W(-n)$ is $\rho^{-8}$, so it does not affect the leading term. The coefficient $\alpha$ was obtained by expanding $S_c$ in the complex coordinate $z=p_1+\iu p_2$. A term of the form $z^{j_1}\bar z^{j_2}/(z\bar z)^{j_3}$ is nonanalytic when $j_1<j_3$ or $j_2<j_3$, with homogeneous degree $j_1+j_2-2j_3$ and angular order $j_1-j_2$. For d3, $(\alpha,\lambda)=(-1/8,1)$; for d1, $(-1/32,3)$; and for d2, $(1/16,3)$; these give the amplitudes in Eq.~(\ref{eq:tail}). For d3, cubic symmetry alone does not forbid the $\lambda=1$ term $|p|e^{-3\iu\varphi_p}=\bar z^2/z$. However, the two contributions to $\delta S_c=\delta\kappa_c/D-S_c\,\delta D/D$ ($\kappa_c=\kappa_1+\iu\kappa_2$) give $\bar z^2/(8z)$ and $-\bar z^2/(8z)$, respectively, and cancel, so that only $-z^3/(8\bar z^2)=-|p|e^{5\iu\varphi_p}/8$ remains at $\lambda=1$. Also for a general $p^3$ deformation, the part proportional to $p_\mu p^2$ produces only analytic terms, so only one coefficient remains at $\lambda=1$. For general local deformations, the amplitudes of the tail all vanish only in the subspace (of codimension at most 3) where the $\lambda=1$ coefficient and the coefficients of the $\lambda=3$ terms with angular orders 5 and 9 all vanish. The contact-term direction (\ref{eq:contact}) lies in this subspace.

The eight-dimensional space of Sec.~IV D is spanned by five deformations that change $w$ by $\sum_\mu\sin^{2j}(p_\mu/2)$ ($j=1$--$4$) and by $\sin^2(p_1/2)\sin^2(p_2/2)$, and three deformations that change $\kappa_\mu$ by $\sin p_\mu\sin^2(p_\mu/2)$, $\sin p_\mu\sin^4(p_\mu/2)$, and $\sin p_\mu\sin^2(p_\nu/2)$ ($\nu\neq\mu$), and it contains (\ref{eq:contact}). The seven nonzero eigenvalues of $\CovB(\Pi^{\rm lat}_\delta,\Pi^{\rm lat}_{\delta'})$ are at least $2\times10^{-8}$ (the basis is not normalized).

\section{Curvature}
\label{app:curv}
Limit to the warped product. We collect the coordinates other than $\sigma_c$ ($\theta_1$ and the sizes $\tau$ of the deformations of the kinetic term) as $y^i$ and write the metric (\ref{eq:metric}) as $(NG^{-1}(0)+F_{00})d\sigma_c^2+2F_{0i}\,d\sigma_c\,dy^i+F_{ij}dy^idy^j$ ($F$ are the components of the covariance part). Rewriting $\sigma_c=\sigma_c^{(0)}+y^0/\sqrt N$ around a reference point $\sigma_c^{(0)}$, the coefficients and their derivatives up to second order converge as $N\to\infty$ to those of the warped product $G^{-1}(0)\,(dy^0)^2+F_{ij}dy^idy^j$, and the sectional curvature is $K=K_\infty+O(1/N)$. $K_\infty$ is the sectional curvature of $F_{ij}$ for planes tangent to the submanifold with constant $\sigma_c$, and $-{\rm Hess}_F\big(G^{-1}(0)^{1/2}\big)(U,U)\big/\big(G^{-1}(0)^{1/2}F(U,U)\big)$ for the plane spanned by $\partial_{y^0}$ and a tangent vector $U$~\cite{ONeill}.

Derivation of Eq.~(\ref{eq:KT}). Taking the coordinates $(\sigma_c,\theta_1)$ and labeling their components by $c$ and $1$, the stationarity condition gives $\theta=(\theta_1,\,2\theta_1\sigma_c-J(\sigma_c))$, and per site $N_s^{-1}\log Z=N\Phi_0+\Phi_1+O(1/N)$, with $\Phi_0=-V(\sigma_c)=\theta_1\sigma_c^2-\sigma_cJ(\sigma_c)+I(\sigma_c)$ and $\Phi_1=-\frac12\bzint\log G^{-1}$. Since $\theta$ is linear in $\theta_1$ at fixed $\sigma_c$, $\partial_{\theta_1}|_{\sigma_c}$ (that is, the direction of $v_m$) is an e-geodesic, and $T_{111}=\partial_{\theta_1}^3(N_s^{-1}\log Z)|_{\sigma_c}=-8\bzint G^3$ ($\Phi_0$ is linear in $\theta_1$). Since the derivative of $\partial_{\sigma_c}\theta$ with respect to $\theta_1$ is $(2/G^{-1}(0))\partial_{\sigma_c}\theta$, $T_{cc1}=\partial_{\theta_1}g_{cc}-4g_{cc}/G^{-1}(0)$, and from $g_{cc}=NG^{-1}(0)+O(1)$ and $\partial_{\theta_1}G^{-1}(0)=2$, $T_{cc1}=-2N+O(1)$. In the curvature of a dually flat space, $R_{c1c1}=-\frac14g^{ij}(T_{cci}T_{11j}-T_{c1i}T_{c1j})$ (whose sign we checked with $K=-1/2$ for the family of one-dimensional Gaussian distributions), only $g^{11}T_{cc1}T_{111}$ and $g^{cc}T_{cc1}^2$ remain at order $N$, and from $\det g=NG^{-1}(0)g_{11}$ and $g_{11}=2\bzint G^2$ we obtain $K_\infty=G(0)T_{111}/(2g_{11}^2)+G(0)^2/g_{11}$. This equals Eq.~(\ref{eq:K}), and when written with $v_\Lambda=\partial_{\sigma_c}$ and $v_m=\frac12\partial_{\theta_1}$, it becomes Eq.~(\ref{eq:KT}). It also agrees with the $N\to\infty$ limit of the Gauss curvature of the Hessian metric computed directly in a model reduced to three modes.

Derivation of Eqs.~(\ref{eq:K}) and (\ref{eq:Kasym}). Inserting $G^{-1}(0)$ ($\partial_{\theta_1}G^{-1}(0)=2$), $F_{\theta_1\theta_1}=2\bzint G^2$, and $\partial_{\theta_1}F_{\theta_1\theta_1}=-8\bzint G^3$ into the formula above gives Eq.~(\ref{eq:K}). Writing $\varepsilon(q)=G^{-1}(q)-1/\glat^2$, Eq.~(\ref{eq:Gbound}) shows that $G^{-1}$ is bounded below by a positive constant and that $|\varepsilon|\le C(1+|\log\hat q|)$, so the Brillouin-zone averages of all powers of $\varepsilon$ are bounded. Expanding $G^2$ and $G^3$ around $1/\glat^2$ to second order in $\varepsilon$ and using $\bzint\varepsilon=0$, we obtain $\bzint G^2=\glat^4[1+3\glat^4\bzint\varepsilon^2+O(\glat^6)]$ and $\bzint G^3\big/\bzint G^2=\glat^2[1+O(\glat^4)]$, which give the first form of Eq.~(\ref{eq:Kasym}). Inserting $G^{-1}(0)=1/\pi+O(a\Lambda\ell)$ and $1/\glat^2=(\ell+\gB)/\pi+O(a\Lambda\ell)$ and replacing $\VarB(\Pi)$ by its massless value gives the second form. The errors of these replacements are $O(a\Lambda\ell^3)$, which is smaller than $O(1/\ell)$ at large $\ell$. In Fig.~\ref{fig:K}, the deviation from $c_K$ at small $\ell$ is due to these $O(a\Lambda\ell^3)$ and $O(1/\ell)$ corrections. The distance follows from $\frac12\bzint(d\log G^{-1}/d\ell)^2=[1+o(1)]/(2(\ell+\gB)^2)$ for the tangent vector $d/d\ell$ of the line. The metric $2\bzint G^2$ in the $\theta_1$ direction is bounded below by $\approx1/(2\theta_1^2)$ independently of $\sigma_c$, since $G\approx1/(2\theta_1)$ at the lattice scale. Numerically, $\gB$ was obtained from $\pi/\glat^2-\ell$ ($0.37144$--$0.37150$ for $a\Lambda=3.9\times10^{-5}$ to $2.4\times10^{-6}$), and $\VarB(\Pi_{\rm lat})$ from the position-space calculation of Appendix C.

Planes including kinetic-term directions (Sec.~V B). The metric of the submanifold with constant $\sigma_c$ is the pullback by the map $H=\log G^{-1}$ into the flat space of functions on the Brillouin zone, so the curvature of the submanifold can be written, for two tangent directions $v$ and $v'$, by the Gauss equation $K(v,v')=[(\Pperp H_{vv},\Pperp H_{v'v'})-|\Pperp H_{vv'}|^2]/(F_{vv}F_{v'v'}-F_{vv'}^2)$ ($(f,f')=\frac12\bzint ff'$ for functions $f$ and $f'$ on the Brillouin zone, $|f|^2=(f,f)$, and $\Pperp$ is the projection onto the component orthogonal to the tangent space). Denoting derivatives by subscripts, $H_{\theta_1}=2G$, $H_\delta=G\Pi^{\rm lat}_\delta$, $H_{\theta_1\theta_1}=-2GH_{\theta_1}$, $H_{\theta_1\delta}=-2GH_\delta$, and $H_{\delta\delta'}=G\Pi^{\rm lat}_{\delta\delta'}-H_\delta H_{\delta'}$ ($\Pi^{\rm lat}_\delta$ and $\Pi^{\rm lat}_{\delta\delta'}$ are the first and second derivatives of the massless self-energy with respect to the deformations of the kinetic term; momentum-independent parts do not affect the curvature). Expanding $G=\glat^2-\glat^4\varepsilon+\glat^4\varepsilon^2G$, we obtain the following. In the plane of the mass-term direction and a kinetic-term direction, $K\to2\big[\bzint(\Pperp\Pi_{\rm lat})\Pi^{\rm lat}_{\delta\delta}\big]/\VarB(\Pi^{\rm lat}_\delta)+O(1/\ell)$ ($\Pperp$ is the projection onto the component orthogonal to $\{1,\Pi^{\rm lat}_\delta\}$), whose limit is finite and equals $-0.2865$ for d1 (from the position-space calculation of Appendix C). If two kinetic-term directions $\delta$ and $\delta'$ are each taken as directions in which $\theta_1$ is moved simultaneously so as not to change $1/\glat^2$, then in their plane $K=(2/\glat^4)K_s[1+O(1/\ell)]=(2/\pi^2)\ell^2K_s[1+O(1/\ell)]$. Here $K_s$ is the Gauss curvature of $\CovB(\Pi^{\rm lat}_\delta,\Pi^{\rm lat}_{\delta'})$ regarded as a metric. In the plane spanned by the $\sigma_c$ direction and a kinetic-term direction in which $\theta_1$ is moved simultaneously so as not to change $G^{-1}(0)$ (which agrees with $v_\delta$ on the line of constant physics up to $O(a\Lambda\ell)$), the warped-product formula above gives $K=-(G(0)/\glat^4)\,Q_\delta\,[1+O(1/\ell)]$, where $Q_\delta$ is the following finite value:
\begin{equation*}
Q_\delta=\frac{\Pi^{\rm lat}_{\delta\delta}(0)-\bzint\Pi^{\rm lat}_{\delta\delta}-\beta\big(\Pi^{\rm lat}_\delta(0)-\bzint\Pi^{\rm lat}_\delta\big)}{\VarB(\Pi^{\rm lat}_\delta)+\big(\Pi^{\rm lat}_\delta(0)-\bzint\Pi^{\rm lat}_\delta\big)^2},\qquad\beta=\frac{\CovB(\Pi^{\rm lat}_{\delta\delta},\Pi^{\rm lat}_\delta)}{\VarB(\Pi^{\rm lat}_\delta)}.
\end{equation*}

Derivation for curvature and phase transitions (Sec.~V C). The jump follows from $\sigma_-=-a\Lambda[1+O(a\Lambda\ell)]$ in Appendix A. The curvature is finite because $G(0)\le C$ and $\bzint G^2>0$ in Eq.~(\ref{eq:K}), and since in each phase the saddle point satisfies $V''(\sigma_c)>0$ and moves smoothly, the curvature stays finite up to the line from one side. The divergence at the endpoint follows from Eq.~(\ref{eq:Kasym}), and the correlation length is determined by the two-fermion threshold $2a\Lambda$. The finiteness of the metric follows from $G^{-1}(0)\to1/\pi$ and $\bzint G^2=\glat^4[1+O(\glat^4)]$. The fraction of the singular part follows from the fact that the contribution of the physical-scale modes ($\hat q\le Ca\Lambda$) to $\bzint G^2$ is at most $C(a\Lambda)^2$, only a fraction $O((a\Lambda)^2\ell^2)$ of the whole.

\section{Limit of the long-distance modes and the momentum-subtraction condition}
\label{app:closed}
Derivation of Eq.~(\ref{eq:closed}). For $|q|\le Ca\Lambda$ and $\sigmabar=a\Lambda$, $\partial_\sigmabar[\Pi(q;\sigmabar)-\Pi(0;\sigmabar)]=\partial_\sigmabar\Delta\Pi_c(q;\sigmabar)+O(1)$ holds (in the small-$q$ estimate of Appendix B, use $\partial_\sigmabar f$ of degree $-3$ instead of $f$). From Eq.~(\ref{eq:Jp}), $a\Lambda\,\partial_\sigmabar\Pi(0;\sigmabar)|_{\sigmabar=a\Lambda}=1/\pi+O(a\Lambda\ell)$. Since $\Delta\Pi_c(q;\sigmabar)$ is a function of $q/\sigmabar$ only and becomes a function of $k/\Lambda$ at $\sigmabar=a\Lambda$, $a\Lambda\,\partial_\sigmabar\Delta\Pi_c=-k\,\partial_k\Delta\Pi_c$. Hence $a\Lambda\,\partial_\sigmabar\Pi(q;\sigmabar)=1/\pi-k\,\partial_k\Delta\Pi_c+O(a\Lambda\ell)$. With $u={\rm arcsinh}(k/(2\Lambda))$, $G_c^{-1}=(1/\pi)u\coth u$, and using $k\,\partial_k=\tanh u\,\partial_u$, $k\,\partial_k\Delta\Pi_c=(1/\pi)(1-2u/\sinh2u)$, which gives Eq.~(\ref{eq:closed}). Since $\chi$ is strictly monotonically decreasing, the cotangent vectors $(\chi(k_i/\Lambda),1)$ of two modes with $|k_1|\neq|k_2|$ are linearly independent, and the rank is 2.

That $\{F^c_k\}$ determines $G_c$. $F^c_k$ is $\frac12(d\log G_c)^2$ of the Gaussian distribution, and since $G_c$ is a function of $k/\Lambda$ only, varying $\Lambda$ reaches all values of $k/\Lambda$, so that $\{F^c_k\}$ determines $d\log G_c/d\log k$. From Eq.~(\ref{eq:closed}), this does not vanish for $k>0$, so $\{F^c_k\}$ determines $G_c$ up to a constant factor (the normalization of the field).

For the momentum-subtraction condition, $G^{-1}(ak_R)=G_c^{-1}(k_R/\Lambda)+O(a\Lambda\ell)$ and the strict monotonicity of $G_c^{-1}$ determine the $\Lambda$ that satisfies $G_c^{-1}(k_R/\Lambda)=c_R$ uniquely, up to relative $O(a\Lambda\ell)$ corrections. That is, the points that satisfy this condition are the same as the points that satisfy $\langle\sigma_n\rangle=a\Lambda$ with that $\Lambda$.

\section{Second derivative of the family of fermionic actions}
\label{app:neg}
We regard the fermionic partition function $Z_F=\int d\bar\psi\,d\psi\,e^{-S}$ as a function of $(m_0,g^2)$ and of the size $\tau$ of the deformation of the kinetic term. $N_s$ is the number of lattice sites. Introducing the auxiliary field and integrating out the fermions, $\log Z_F/(NN_s)=-V(\sigma_c)-m_0^2/(2g^2)+O(1/N)$. With $\sigmabar$ fixed, $V$ depends on $(m_0,g^2)$ through $\theta_1$ and $\theta_2$, and on $\tau$ through $I$. The second derivative including the motion of the saddle point is $\partial_\tau^2[-V(\sigma_c)]=\partial_\tau^2I+(\partial_\sigmabar\partial_\tau I)^2/\partial_\sigmabar^2V$, with $\partial_\sigmabar^2V=1/g^2-\partial_\sigmabar^2I=1/g^2+\Pi(0)=G^{-1}(0)$. For deformations that change only $w$ ($\delta\kappa=0$), $D$ depends on $\tau$ only through $M$, so at leading order in large $N$,
\begin{equation}
\frac1{NN_s}\partial_\tau^2\log Z_F=\bzint\omega\,\delta w^2+G(0)\big(\bzint\omega\,\delta w\big)^2,\qquad \omega=\partial_M^2\log D=\frac{2(\kappa^2-M^2)}{D^2}.
\label{eq:negform}
\end{equation}
The second term is a nonnegative term that comes from the motion of the saddle point. The integrand of the first term is negative around the doublers, because there the mass $M$ due to the Wilson term is larger than $|\kappa|$ and $\log D$ is concave in $M$. Equation~(\ref{eq:negform}) equals $-0.619$ for d1 and $-5.90$ for d2 (values in the $a\to0$ limit, that is, at $\sigmabar\to0$ and $G(0)=\pi$; on the line of constant physics at $a\Lambda=0.01$, $-0.627$ and $-6.12$). Hence the second derivative of $\log Z_F$ is not positive definite. Since the weight $e^{-S}$ of Grassmann variables is not a probability distribution, the connected two-point functions of the operators that multiply the couplings are not necessarily nonnegative.

\end{document}